\documentclass[a4paper, superscriptaddress,groupedaddress,nofootnoteinbib,11pt]{article}  
\pdfoutput=1
\usepackage{amssymb,amsmath,graphicx,color,microtype}
\usepackage{varioref}
\usepackage{graphicx}
\usepackage{txfonts}
\usepackage{jcappub}
\usepackage{epsf}
\usepackage{multirow}
\usepackage{epstopdf}
\usepackage {amssymb}
\usepackage{tcolorbox}
\usepackage{tabularx}
\usepackage{array}
\usepackage{colortbl}
\tcbuselibrary{skins}
\usepackage{booktabs}

\newcolumntype{Y}{>{\raggedleft\arraybackslash}X}
\newcommand{\nc}{\newcommand}
\nc{\ba}{\begin{eqnarray}}
\nc{\ea}{\end{eqnarray}}
\newcommand\be{\begin{equation}}
\newcommand\ee{\end{equation}}

\usepackage{tabularx}
\tcbset{tab2/.style={enhanced,fonttitle=\bfseries,fontupper=\normalsize\sffamily,
colback=yellow!10!white,colframe=red!50!black,colbacktitle=Salmon!40!white,
coltitle=black,center title}}

\nc{\e}{{\bf{e}}}
\nc{\kk}{{\bf{k}}}
\nc{\pp}{{\bf{p}}}

\nc{\bfk}{{\bf{k}}}
\nc{\bfx}{{\bf{x}}}
\nc{\bfp}{{\bf{p}}}

\nc{\eH}{{\epsilon_H}}
\nc{\calP}{{\cal P}}
\nc{\im}{{ \mathrm{Im} } }

\begin{document}

%%%%%%%%%%%%%%%%%%%%%%%%%%%%%%%%%%%%%%%%%%%%%%%%%%%%%
\title{Observational signatures of the black hole Mass Distribution in the Galactic Center}

\author{Razieh~Emami and Abraham~Loeb }
%\email{razieh.emami_meibody@cfa.harvard.edu}
%\email{aloeb@cfa.harvard.edu}
\affiliation{Center for Astrophysics, Harvard-Smithsonian, 60 Garden Street, Cambridge, MA 02138, USA}

\abstract{\\ 
We simulate the star cluster, made of stars in the main sequence and different black hole (BH) remnants, around SgrA* at the center of the Milky Way galaxy. Tracking stellar evolution, we find the BH remnant masses and construct the BH mass function. We sample 4 BH species and consider the impact of the mass-function in the dynamical evolution of system. Starting from an initial 6 dimensional family of parameters and using an MCMC approach, we find the best fits to various parameters of model by directly comparing the results of the simulations after $t = 10.5$ \rm{Gyrs} with current observations of the stellar surface density, stellar mass profile and the mass of SgrA*. Using these parameters, we study the dynamical evolution of system in detail. We also explore the mass-growth of SgrA* due to tidally disrupted stars and swallowed BHs. We show that the consumed mass is dominated for the BH component with larger initial normalization as given by the BH mass-function. Assuming that about 10\% of the tidally disrupted stars contribute in the growth of SgrA* mass, stars make up the second dominant effect in enhancing the mass of SgrA*. We consider the detectability of the GW signal from inspiralling stellar mass BHs around SgrA* with LISA. Computing the fraction of the lifetime of every BH species in the LISA band, with signal to noise ratio $\gtrsim 8$, to their entire lifetime, and rescaling this number with the total number of BHs in the system, we find that the total expected rate of inspirals per Milky-Way sized galaxy per year is $10^{-5}$. Quite interestingly,   
the rate is dominated for the BH component with larger initial normalization as dictated by the BH mass-function. 
We interpret it as the second signature of the BH mass-function.
}

\maketitle
%%%%%%%%%%%%%%%%%%%%%%%%%%%%%%%%%%%%%%%%%%%%%%%%%%%%%
\section{Introduction}

The Supermassive black hole (SMBH) at the Galactic center, Sgr A*, is surrounded by a 
cluster of stars and their remnant black holes (BHs), which migrate inwards by 
dynamical friction. An exhaustive consideration of this process was performed in Refs. \cite{Bahcall-Wolf-1976, Freitag:2006qf, Alexander:2008tq} which considered 
the formation of a steep density cusp with high velocity dispersion around the galactic center.
Mass segregation of heavier objects concentrates them around the SMBH. This generates 
gravitational waves (GWs) which are observable at cosmological distances
\cite{Hopman:2006xn, Alexander:2008tq,Hopman:2005vr,Preto:2009kd,Alexander:2017arl,Babak:2017tow}.
In addition, the close encounters between the stellar mass BHs can also create GWs sources
\cite{OLeary:2008myb, Banerjee:2009hs, Antonini:2012ad, Boehle,Berry:2013ara, Antonini:2016gqe,  Aharon:2016kil}.
Another observational signature of dense stellar cluster around galactic centers is X-ray binaries due to the tidal capture of stars by BHs and neutron stars (NSs). This was recently investigated in Ref. \cite{Generozov:2018niv} and applied to 
the $\gamma$ ray excess at the center of the  Milky Way (MW) \cite{Freitag:2006qf, Auchettl:2016qfa, Arca-Sedda:2017qcq,Perna:2019axr,Bortolas:2017moe, Generozov:2018niv}.
A further interesting observable may originate from mergers of BHs and neutron stars at the Galactic center \cite{Fragione:2018yrb, Chatterjee:2001ty, Chatterjee:2002bg}.

There are various ways to model the behavior of such a system. N-body and Monte Carlo codes are computationally very expensive. Using the distribution function is favoured computationally for  describing different elements and finding the time evolution of the entire system at different time snapshots. The evolution of the system is given by the Fokker Planck equation, coupled to the Poisson equation. In Ref. \cite{Vasiliev:2017sbo},  the Fokker Planck approach was carefully studied confirming explicitly that the results from this approach are  well matched with the results from numerical simulations.

Here we simulate the star cluster at the center of the Milky Way galaxy using the Fokker Planck approach presented in the
Phase Flow, hereafter (PF), code which is a library in the publicly available code AGAMA \footnote{https://github.com/GalacticDynamics-Oxford/Agama} \cite{Vasiliev:2017sbo}. Our system is made of a SMBH, Sgr A*, main sequence stars and four species of the stellar-mass BHs as the remnant to the main sequence stars. For the first time, in this context, we use the publicly available tool COSMIC \footnote{https://github.com/COSMIC-PopSynth/COSMIC} 
(Compact Object Synthesis and Monte Carlo Investigation Code) to find the BH remnant mass as a function of the zero age main sequence (ZAMS) mass focusing only on $Z = 0.001$. Work is in progress to consider the impact of different metallicities in the analysis as well as enlarging the populations of stellar-mass BHs to more BHs. We consider the continuous star formation in our setup as well. We study a six dimensional parameter space, as our initial condition, and evolve the system for a total period of $t = 10.5$ \rm{Gyrs}. We find the best fit to these parameters by directly comparing the results after $t = 10.5$ \rm{Gyrs} with the most recent observations. Using the best fit values, we study the evolution of the stellar cluster in details. For the first time, we directly consider the impact of the BH mass-function in the simulation of PF. One of the most important places that the impact of BH mass-function directly appears is in the consumed mass to SgrA* from different BH species where we find that the contribution from the second heaviest BH is dominated over the rest of the BHs as well as the stars in the main sequence. Assuming that 10\% of the disrupted stars contribute in the growth of the central BH mass, we find that stars are the second most dominant contributor in the mass growth of SgrA*.

As another application of this approach, we study the GW from the inspiral phase of stellar mass BHs around SgrA*. Starting from a sample of 10000 systems, we consider dynamical evolution of the semi-major axes as well as the eccentricity including the impact of the angular momentum diffusion. We compute the fraction of the lifetime of every systems in the LISA band with the signal to noise ratio of 8 or above to their actual lifetime and use this as a weighting factor in our estimation. Furthermore, we also rescale this with the actual number of the BHs interior to $r = 10 pc$ as the upper limit of our sampling. We compute the expected rate of inspiral for the Milky Way (MW) like galaxies. Our results are in good agreement with the inferred inspiral rate computed from totally other methods.

Our simulations  are based on various assumptions and simplifications as listed below:

$(1)$  We assume spherical symmetry for the MW core  \cite{Levin:2003kp}. 

$(2)$ We neglect gas inflow in our simulation, the modification of stellar orbits through
hydrodynamical drag \cite{Kennedy:2016tyr}  and through non-spherical contributions (e.g. disk like) to the global gravitational potential \cite{Karas:2007ds}.

$(3)$ We neglect resonant relaxation. This includes both of scalar and vector resonances as considered in some details at \cite{Merritt:2015vxa, Merritt:2015xpa,Merritt:2015elb, Merritt:2015kba}.
The importance of scalar resonant relaxation (RR) is still under debate, with controversy in the literature about its importance. While Ref. \cite{Hopman:2006qr}  pointed out that the scalar RR affects the tightly bound stars and enhances the tidal disruption rates, the detail consideration in Ref. \cite{Merritt:2011ve} showed that relativistic precession suppresses RR and so it is not very efficient. 
This conclusion is consistent with Ref. \cite{Bar-Or:2016qop} where the authors showed that the effect of  scalar RR is small (see Fig. 17 of this paper). Vector RR, on the other hand, is harder to predict and it is not obvious whether it might have important effects on the evolution of the system. If important, it would affect the spherical symmetry of the system \cite{Szolgyen:2018zra}. 

$(4)$ Another neglected effect is close binaries near SMBH. As discussed briefly in \cite{Bahcall-Wolf-1977} the impact of binaries on the cusp dynamics is small. As the vast majority of these binaries are soft the total transferred energy by them to the cusp is small.  Dynamical evolution of binaries interacting with the stars around SMBH was further performed in Ref. \cite{Hopman:2009gz} where the author showed that most of the binaries are disrupted either by the loss-cone effects, happening far away from the SMBH, or through the three body encounters, very close to the center. The main conclusion was that binaries are disrupted before they could experience an exchange interaction. 
Although very close to the GC secular Kozai evolution affects the dynamics of binaries
and  periodically change their inclinations and eccentricities, these populations are continuously destroyed by different processes such as the evaporation, mergers and disruption \cite{Wen:2002km,Kocsis:2011jy, Antonini:2013tea,VanLandingham:2016ccd, Hoang:2019kye}. An extensive study of the binary population around SMBH including all of these effects is beyond the scope of this paper and is left for a future work.

The outline of our paper is as follows. In section \ref{Fokker-Planck} we review the Fokker Planck approach. In section \ref{initial-profile} we describe the initial conditions of the simulation. Using an MCMC approach, we find the best fit for a 6 dimensional family of parameters and use them as our initial conditions throughout our simulations. 
In section \ref{Relaxation} we study the two body relaxation in detail and consider the time evolution of Sgr A*, as well as the mass profile of the stars and BHs. We also compute the mass-radius relation for the BHs at a few different times and initial slopes. In section \ref{GW}, we consider the detectability of this system with LISA . We conclude in section \ref{conclusion}. 

%%%%%%%%%%%%%%%%%%%%%%%%%%%%%%%%%%%%%%%%%%%%%%%%%%%%%
\section{Fokker-Planck approach for the relaxation process}
\label{Fokker-Planck}
Here we describe our simulation setup. We use the PF code which is based on a Fokker Planck approach. We compute the time evolution of the distribution function of a muti-body system made of the stars on the main sequence and different BH species being influenced by a central super-massive BH. 

\ba
\label{Fokker-Planck1}
\frac{\partial f_c (h,t)}{\partial t} = - \frac{\partial \textit{F}_c(h,t)}{\partial h} - \nu_c(h,t) f_{c}(h,t) + S(t),
\ea
here $f_c$ refers to the phase space distribution function of every species, with sub-index $c$ referring to stars and BHs. Eq. (\ref{Fokker-Planck1}) contains a few additional parameters, $h(E), \textit{F}_c(h, t)$, $\nu_c(h, t)$  and $S(t)$. 

In the following, we provide a brief introduction to each of these components,

$\bullet$ \textbf{Phase space volume, $h(E)$:} refers to the 
volume enclosed by an energy hypersurface $E$ \cite{Vasiliev:2017sbo},  
\ba 
\label{phase-space-volume}
h(E) = 16 \pi^2 \int_{\Phi(0)}^{E} dE' \int_{0}^{r_{max}(E')} r^2 \sqrt{2\left(E' - \Phi(r) \right)} dr,
\ea
where $\Phi(r)$ describes the total gravitational potential, 
\ba 
\label{Phi}
\Phi(r) &=& - \frac{G M_{\bullet}}{r} - 4 \pi G \sum_{i} \bigg{[} \left( \frac{1}{r}\right) \int_{0}^{r} dr' r'^2 \rho_i(r')  
+ \int_{r}^{\infty}  dr' r'  \rho_i(r') \bigg{]}.
\ea
and $\rho_i(r) = 4 \pi  \int_{h(\Phi(r))}^{\infty} dh' \frac{f_i(h')}{g(h')} \sqrt{2 [E(h') - \Phi(r)]}$ refers to the mass density for every species. Finally, $g(h) \equiv \frac{d h (E)}{dE}$ denotes the density of states \cite{Vasiliev:2017sbo}. 

$\bullet$ \textbf{ Mass Flux, $\textit{F}_c(h,t)$:} 
describes the mass flux through $h$ and is given by, 
\ba 
\label{mass-flux}
\textit{F}_c(h,t) = A_c f_c + D \frac{\partial f_c}{\partial h},
\ea
where $A_c(h)$, $D(h)$ refer to the advection and diffusion coefficient, respectively, 

\ba 
\label{advection}
A_c(h)  &=& 16 \pi^2 G^2 \ln{\Lambda} ~ m_c \sum_i \int_{0}^{h} f_i(h') dh', \\
\label{diffusion}
D(h) &=& 16 \pi^2 G^2 \ln{\Lambda}~ g(h) \sum_i  m_i \bigg{(} \int_{0}^{h} \frac{f_i(h') h'}{g(h')} dh' + 
h  \int_{h}^{\infty} \frac{f_i(h')}{g(h')} dh' \bigg{)}.
\ea
The summation in Eqs. (\ref{advection}) and (\ref{diffusion}) is over different species. While the advection term is proportional to the mass of individual species, the diffusion coefficient is exactly the same for all of them. In our analysis, we adopt a Coulomb logarithm \cite{Binney-Tremaine} $\ln{\Lambda} \simeq \ln{ \left( \frac{M_{\bullet}}{M_{*}} \right)}\simeq 10$. 

$\bullet$ \textbf{ Sink term, $\nu_c(h,t)$:} 
SMBH acts like a sink as it consumes stars and compact objects which arrive sufficiently close to it. This happens in one of the two ways. Stars get disrupted by the tidal force of the SMBH at the distance $R_{tid} \equiv R_{\star} \left(M_{\bullet}/M_{\star}\right)^{1/3} $ where $R_{\star}$ and $M_{\star}$ refer to the radius and the mass of a typical star, respectively. The mass and the radius of the star are correlated, and we use the fitting formula in Ref.  \cite{Rubin:2010bw}.  Not the entire mass of the disrupted star is added to the mass of SMBH. Only a  fraction of its mass contributes in enhancing the SMBH's mass. A new parameter,  $f_{dis}$, is introduced to refer to the fraction of mass of a typical star that would be added to the central BH mass, $f_{dis} M_{\star}$. 
In the following, we consider two different examples with $f_{dis} = (1\%, 10\%)$. 
Compact objects in the form of stellar BHs, on the other hand, are being eaten by the SMBH within the capture radius $R_{cap} = 8 G M_{\bullet}/c^2$. Here we assume that the entire mass of the swallowed BHs are added to the SMBH mass.

In the following, we use $R_{tid}$ and $R_{cap}$ for the so called loss-core radius, hereafter $r_{LC}$. Consuming the neighboring objects by the SMBH lead  to an increase in its mass. Here we follow the approach of Ref. \cite{cohn1987} to model this effect using the one dimensional orbit averaged Fokker-Planck equation. Using this approach, the loss term is estimated as, 

\ba 
\label{loss-rate}
\nu = \frac{\mu(E)}{\alpha + \ln{\left(1/{\mathcal{R}}_{LC}\right)}} ~~~,~~~ \alpha \simeq \left(q ^2 + q^4 \right)^{1/4},
\ea
where we have removed the sub-index $c$ for the sake of simplicity.  $q$ refers to the loss-cone filling factor and is given by \cite{Vasiliev:2017sbo},

\ba 
\label{filling-factor}
q (E) \equiv \frac{\mu(E) P(E)}{ \mathcal{R}_{LC}(E)} ~~,~~ \mathcal{R}_{LC}(E) \equiv \frac{L^2_{LC}}{L^2_{cir}} = \frac{2 G M_{\bullet} r_{LC}}{L^2_{cir}(E)},
\ea
where $\mathcal{R}_{LC}(E) $ refers to the width of the loss-cone and is given by Eq. (\ref{filling-factor}). $L^2_{cir}(E)$ and $L^2_{LC}$ denote the angular momentum of the circular orbits and the loss-cone, respectively. In addition, $P(E)$ refers to the period of nearly radial orbits with energy $E$. 

Finally $\mu(E)$ denotes the orbit averaged diffusion coefficient and is given by,

\ba 
\label{mu}
\mu(E) \equiv \frac{8 \pi^2 }{g(h(E))} \int_{0}^{r_{max}} \frac{\langle \Delta v^2_{\perp}\rangle r^2 dr}{\sqrt{2(E - \Phi(r))}}, 
\ea
where $r_{max}$ refers to the maximum radius  accessible to star with total energy $E$ and is given by $\Phi(r_{max}) = E$. In addition, $\langle \Delta v^2_{\perp}\rangle $ is given by, 

\ba 
\langle \Delta v^2_{\perp}\rangle  &\equiv & 16 \pi^2 G^2 \ln{\Lambda} \sum_{i} m_i \bigg{(} 
\frac{4}{3} \int_{E}^{0} f(E') dE' + 2 \int_{\Phi(r)}^{E} dE' f_{i}(E') \left( 
\frac{E' - \Phi(r)}{E - \Phi(r)}
\right)^{1/2} \nonumber\\
&& ~~~-\frac{2}{3} \int_{\Phi(r)}^{E} dE' f_{i}(E') \left( 
\frac{E' - \Phi(r)}{E - \Phi(r)}
\right)^{3/2}
\bigg{)}. \nonumber\\
\ea 
In Sec. \ref{evolution-cbh-mass} we consider a signature of the above sinking process which is shown as an enhancement of the SMBH mass with time.  

$\bullet$ \textbf{ Source term, $S(t)$:} 
We have also added a source term in the Fokker Planck equation, $S(t)$, which refers to the continuous star formation. Here for simplicity we take this function to be constant in time and we consider the source to happen interior to a given radius with a density profile proportional to $\sqrt{1/r - 1/r_{source}}$. So we add two free parameters in our fittings in the next section as the mass of the source as well as the enclosed radius $r_{source}$. Work is in progress to generalize this to the case with an extended source with varies both in time and radii \cite{Emami2020}.

%%%%%%%%%%%%%%%%%%%%%%%%%%%%%%%%%%%%%%%%%%%%%%%

\section{Initial density profile}
\label{initial-profile}

\subsection{Parameterization of the initial density profile}
One of the key ingredients in the Fokker Planck approach is the initial condition for the density profile of different species in the system. Initial conditions are very important since the subsequent evolution of the system strongly and non-linearly depends on their selection. However, since we are dealing with very long period of evolution we have to be very careful in identifying the initial conditions. The best way is to use some parametric choices for the initial density profile and figure out the values for different parameters by comparing the ``evolved'' profiles with the current constraints in the galactic center. Having this said, in what follows, we first make some selections for the parametric initial conditions. Then, we determine different parameters using the direct comparisons with observations.

Hereafter, we take the following functional form for the initial density profile of different species,

\ba 
\label{spheroidal}
\rho(r) = \rho_0 \left(\frac{r}{R_{scale}}\right)^{-\gamma} \left[1 +\left( \frac{r}{R_{scale}} \right)^{\alpha}\right]^{(\gamma - \beta)/\alpha}.
\ea

There are five different parameters in Eq. (\ref{spheroidal}). Not all of the parameters above are important, though. Here
we fix two of these parameters, $\alpha = 4 $ and $\beta = 5$ and find the rest of them by a direct comparison with observations. We have checked that changing the above values for $\alpha$ and $\beta$ do not affect our results. We are therefore left with three parameters in the density profile. As we will mention, due to computational reasons, in our fitting we replace the density normalization with the mass normalization. This together with $R_{scale}$ and $\gamma$ would be determined later.

%%%%%%%%%%%%%%%%%%%%%%%%%%%%%%%%%%%%%%%%%%%%%%%%
\begin{figure}
\center
\includegraphics[width=0.75\textwidth]{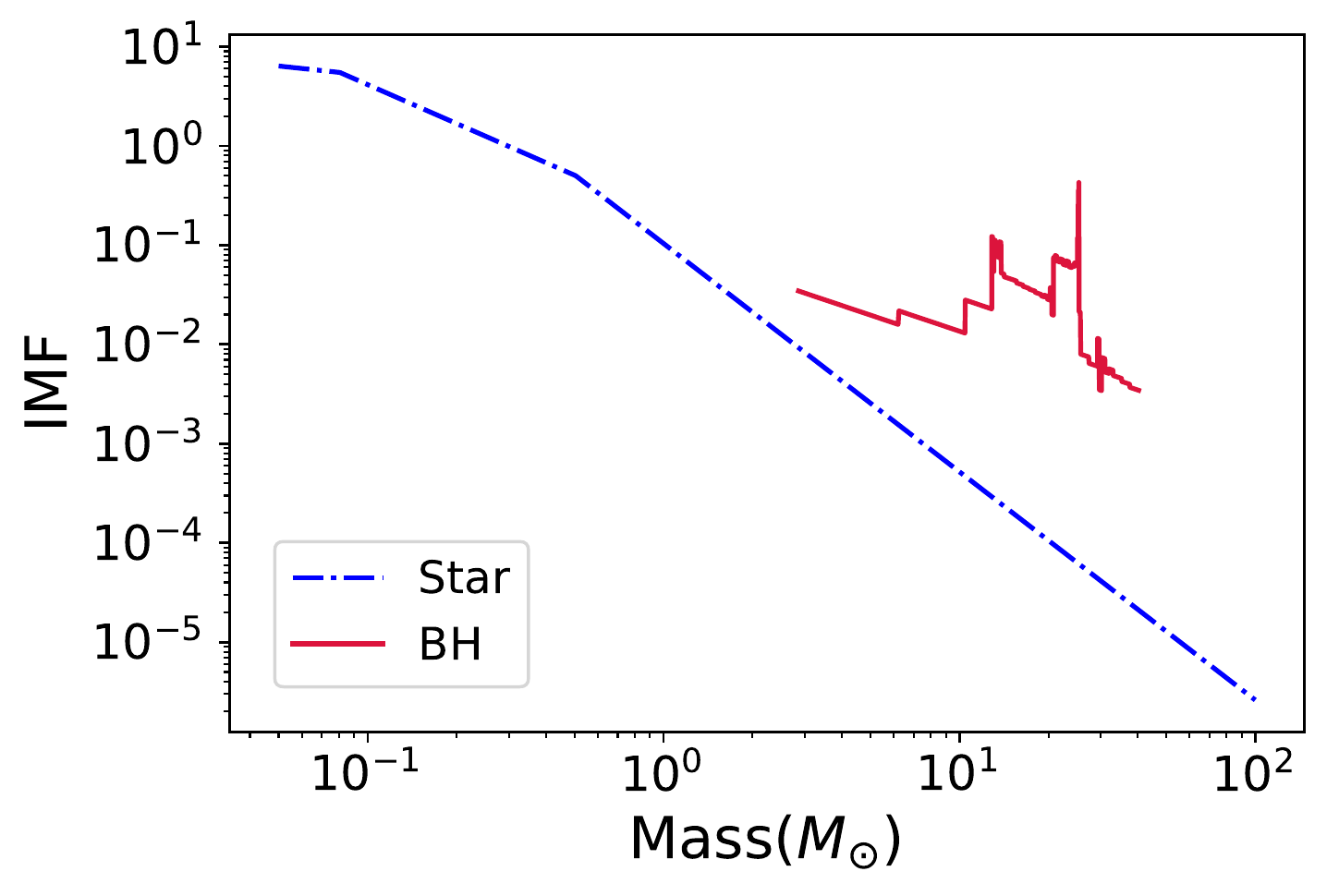}
\caption{Initial mass function (IMF) of the BHs remnants (red line) compared with the initial Kroupa mass function of stars in the ZAMS (dotted dashed blue line). Here we have chosen $Z = 0.001$.}
\label{BH-IMF}
\end{figure}
%%%%%%%%%%%%%%%%%%%%%%%%%%%%%%%%%%%%%%%%%%%%%%%%

While the scaling radius, $a$, and the density slope, $\gamma$, are taken to be the same for different species, the overall normalization in the density, or equivalently in the mass, depends on the mass and abundance of them. Therefore we need to fix the mass and the number of different species in the simulation box. 
We choose 5 different species in the entire of our analysis. One in the form of star in the main sequence with a characteristic mass $m_{\star} = 1 M_{\odot}$. This is due to the fact that heavier stars have shorter lifetimes. Therefore the dynamically dominant component of stars at the present time would be the lower mass stars. We have also considered four BH species with different masses and abundances. 

Using the COSMIC code, we find the BH remnant mass as a function of the zero age main sequence (ZAMS) mass. This strongly depends on the value of the initial metallicity, Z, that we take to be rather small $Z = 0.001$. Using the remnant BH-ZAMS map and by assuming a Kroupa initial mass function (IMF) for the stars, we find the IMF for the BHs. Figure \ref{BH-IMF} presents the IMF of BHs compared with the initial Kroupa mass function of the stars. Using the BH IMF, we make some bins in the mass of BHs and find the mean mass as well as the normalization of them with respect to the stars. This is done in a few steps. First we compute the overall normalization in the entire of BHs using the Kroupa mass function for the stars.

\ba 
\label{BH-norm1}
\rho_{BH,0} &=& \rho_{*,0} \frac{\int_{m_{min}}^{m_{max}} \Phi^{*} m_{rem}(m_*) dm_*}{\int_{m_0}^{m_{max}}\Phi^{*} m_{rem}(m_*) dm_*} = 0.097 \rho_{*,0}.
\ea
where $\Phi^* \equiv dN_{\star}/dM_{\star}$ denotes the Kroupa mass function for the stars and $m_{rem}$ refers to the remnant mass as a function of ZAMS inferred from the COSMIC code. In the numerator, we take  $m_{min} = 18.59 M_{\odot}$
as a threshold to get BH as a remnant (using the COSMIC code). We also limit ourselves to masses less than $m_{max} = 100 M_{\odot}$. The lower limit on the denominator is chosen to be $m_{min} = 0.05 M_{\odot}$ as well.

Next, we use the BH IMF and make some regular bins in the entire range of BH mass to compute the density normalization in a given bin.
Eq. (\ref{Normalization_BH}) presents BH density normalization in bin (i). Here $\Phi_{BH}$ refers to the BH IMF, as shown in Figure \ref{BH-IMF}, $m_i$ and $m_{i+1}$ denote the lower and upper limits of the BH mass in i-th bin and $m_{min, BH} = 2.86 M_{\odot}$ and $m_{max, BH} = 40.35 M_{\odot}$ refer to the minimum and maximum values of the BH mass. 

\ba 
\label{Normalization_BH}
\rho_{BH,0}^{(i)} = \rho_{BH,0} \frac{
\int_{m_i}^{m_{i+1}} \Phi_{BH}(m) m dm} {\int_{m_{min, BH}}^{m_{max,BH}} \Phi_{BH}(m) m dm
}.
\ea

We also compute the mean value of the BH mass in every mass bins. This gives us a consistent link between the BHs mass/normalization and the ZAMS normalization. Table \ref{BH-ZAMS-Norm} presents the mean mass as well as the normalization of the BH densities for our four BH species. We emphasis here that the second column gives us the ``fractional'' normalization in the BH mass as refereed by F[i]. Therefore the actual normalization in the density/Mass of the BHs is given by this number multiplied with the actual normalization of the ZAMS, $M_{\star}$, which is a parameter in our fits. Since the BHs are the remnant of the stars in the ZAMS, we take the normalization of the stars to be $(F[0] = 1- \sum_i F[i]) \times M_{\star}$.

%%%%%%%%%%%%%%%%%%%%%%%%%%%%%%%%%%%%%%%%%%%%%%%%%%%

\begin{table}
\centering
\caption{Map between the BH mass and its fractional normalization in density/Mass compared with the ZAMS normalization}
\label{BH-ZAMS-Norm}
\begin{tabular}{lcr} 
\hline
\hline
Mass($M_{\odot}$)[i] & 
F[i]\\
\hline
8.15
& 6.95 $ \times 10^{-3}$
\\
\hline
16.24
&  3.11 $\times 10^{-2}$
\\
\hline
23.92
& 
5.01 $\times 10^{-2}$
\\
\hline
34.50 &
 7.62 $\times 10^{-3}$
 \\
\hline
\end{tabular}
\end{table}
%%%%%%%%%%%%%%%%%%%%%%%%%%%%%%%%%%%%%%%%%%%%%%%%%%%

In summary, we are left with three free parameters in the density profiles. On the other hand, due to the consumption of the disrupted stars and swallowed BHs, central BH mass increases with the time. Therefore, its initial mass is another free parameter that must be determined as well. Finally, there are two more parameters associated with the source term, i.e. fractional source mass and its radius, that should be found as well. 
Here we split the total mass in every species to two different parts. One is the continuously formed element, $fraC_{source} F[\mu] M_{\star}$, where $fraC_{source}$ refers to the fraction of the continuously born source and $\mu = (0, 1, 2, 3,4) $ refers to both of the stars and different BH species. And, the second part is the initial normalization $\left(1 - fraC_{source}\right) F[\mu] M_{\star}$. 
We name the source radius as $r_{source}$.
This gives us a family of 6 parameters that should be fixed using the observations.

\subsection{Inferring the initial density profile using the observations}
Having introduced a model with 6 free parameters, here we attempt to find each of these parameters by using a direct comparison with observations. We particularly use the data from \cite{Schodel:2017vjf} including the flux density, see the left panel in Figure 9 of \cite{Schodel:2017vjf}, at different radii as well as the enclosed mass at two different radii, namely $ M(r \leq 1pc) = 10^{6} M_{\odot}$ and $ M(r \leq 4pc) = 10^{7} M_{\odot}$. Finally, we assume that after 10.5 \rm{Gyr} the central BH mass is $M_{\bullet} = 4 \times 10^6 M_{\odot}$.
Using the above observations and by performing an MCMC analysis, we find the proper values for the above 6 parameter family. The only left over parameter is the fraction of the tidally disrupted stars added to the central BH mass. Since only a small fraction of the tidally disrupted stars are added to the central BH mass, here consider $f_{dis} = (1\%, 10\%)$ and find the actual fit for both of these choices.
We aim to find some ranges for the density slope, $\gamma$, logarithm of the scaling radius, $\log_{10}(R_\mathrm{scale})$, logarithm of the initial total mass normalization, $\log_{10}(M_\mathrm{*,init})$, logarithm of the initial central BH mass, $\log_{10}(M_\mathrm{\bullet,init})$, fraction of the source mass, $\mathrm{frac}_\mathrm{source}$, and the logarithm of the source radius, $\log_{10}(r_\mathrm{source})$.

%%%%%%%%%%%%%%%%%%%%%%%%%%%%%%%%%%%%%%%%%%%%%%%%
\begin{figure}
\center
\includegraphics[width=1.\textwidth]{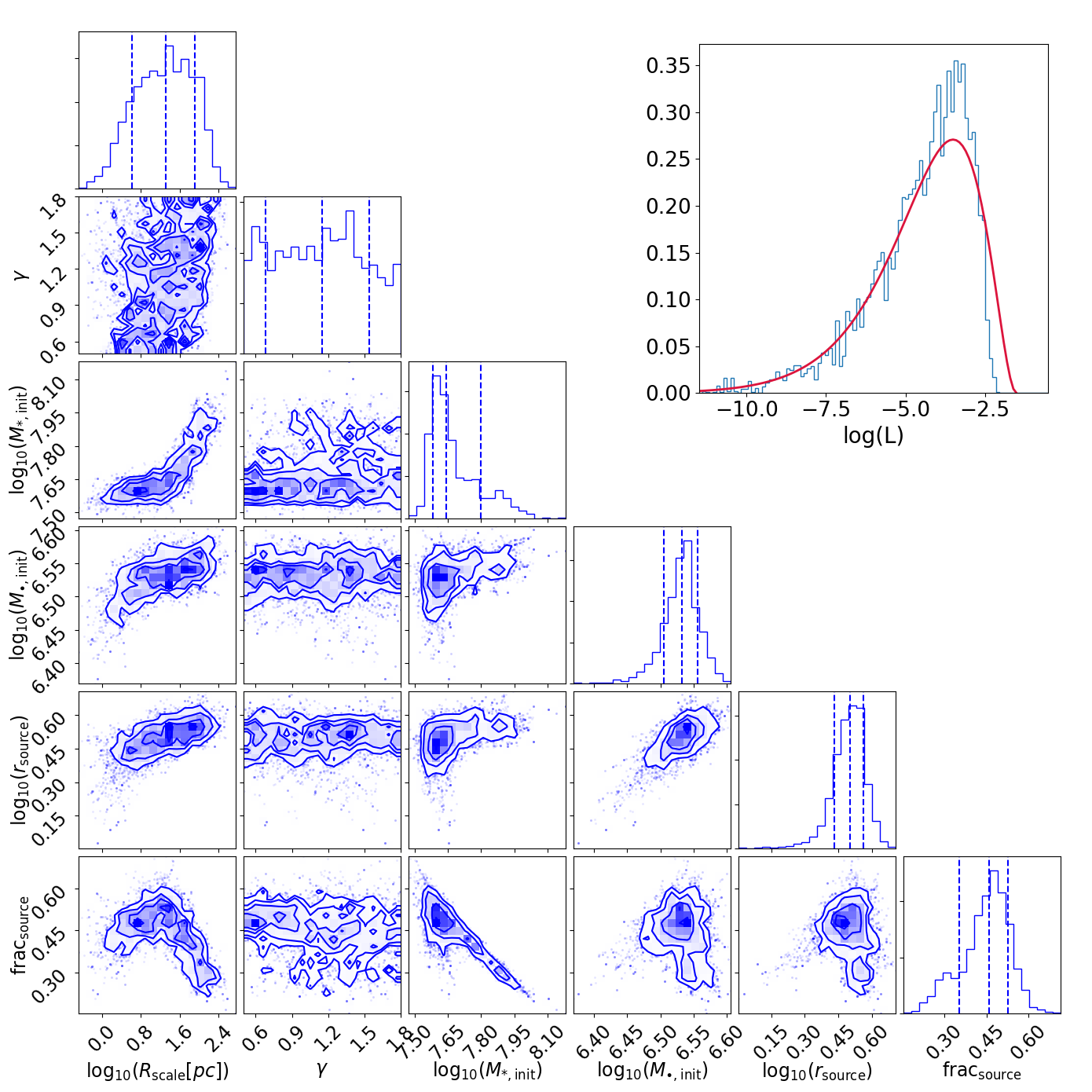}
\caption{Posterior values of the 6 parameter family for the case with $f_{dis} = 1\%$. }
\label{1percent-tidal}
\end{figure}
%%%%%%%%%%%%%%%%%%%%%%%%%%%%%%%%%%%%%%%%%%%%%%%%

 \subsubsection{Parameter fits for $f_{dis} = 1\%$}
Here we find the actual fit to observations when only one percent of the disrupted stars are added to the central BH mass. Figure \ref{1percent-tidal} presents the posterior plot for the range of the parameters. One sigma range of the above 6 dimensional family are given as, 

\ba 
\label{6-dim-1percent}
\gamma &=& 1.14_{-0.43}^{+ 0.46} ~~, \\
\log_{10}(R_\mathrm{scale} [pc]) &=& 1.12_{-0.60}^{+ 0.93} ~~, \\
\log_{10}(M_\mathrm{*,init}) &=& 7.64_{-0.06}^{+0.16} ~~ , \\
\log_{10}(M_\mathrm{\bullet,init})&=& 6.53_{-0.03}^{+0.02} ~~, \\
\mathrm{frac}_\mathrm{source} &=& 0.46_{-0.11}^{+0.07} ~~, \\
\log_{10}(r_\mathrm{source}) &=& 0.50_{-0.07}^{+0.06} ~~.
\ea

Having determined the median and 16, 84 percentile for the above parameters, in the following, we use the median values and we run the Phase-Flow code to find the actual evolution of the system.

 \subsubsection{Parameter fits for $f_{dis} = 10\%$}
\begin{figure}
\center
\includegraphics[width=1.\textwidth]{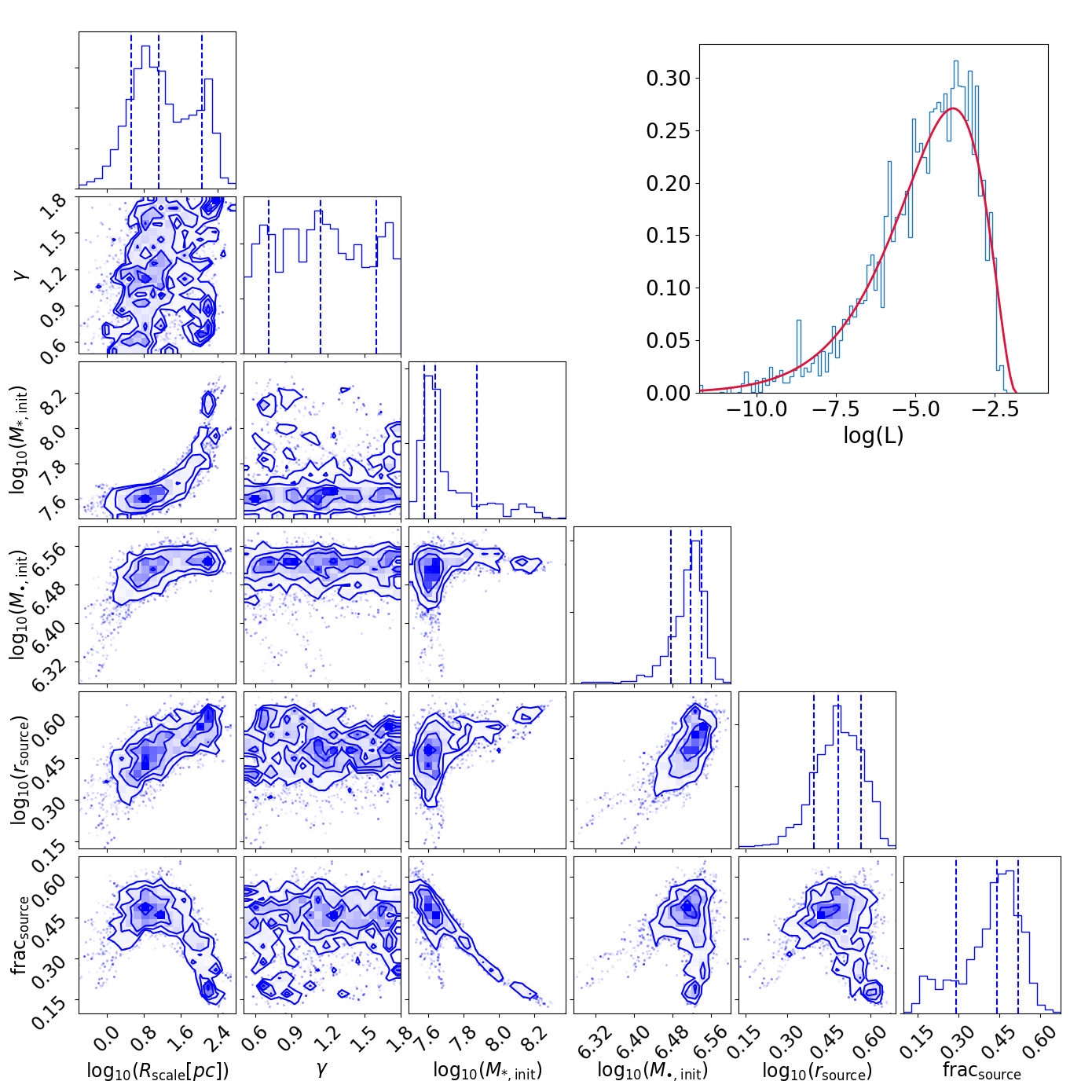}
\caption{Posterior values of the 6 parameter family for the case with $f_{dis} = 10\%$. }
\label{10percent-tidal}
\end{figure}

Here we find the fits for the case with 10\% of the tidally disrupted stars added to the mass of the central BH mass. Figure \ref{10percent-tidal} presents the posterior plot for the range of the parameters. One sigma range of the above 6 dimensional family are given as,

%%%%%%%%%%%%%%%%%%%%%%%%%%%%%%%%%%%%%%%%%%%%%%%%
%%%%%%%%%%%%%%%%%%%%%%%%%%%%%%%%%%%%%%%%%%%%%%%%

\ba 
\label{6-dim-10percent}
\gamma &=& 1.15_{-0.46}^{+ 0.39} ~~, \\
\log_{10}(R_\mathrm{scale} [pc]) &=& 1.31_{-0.70}^{+ 0.6} ~~, \\
\log_{10}(M_\mathrm{*,init}) &=& 7.64_{-0.06}^{+0.23} ~~ , \\
\log_{10}(M_\mathrm{\bullet,init})&=& 6.52_{-0.04}^{+0.02} ~~, \\
\mathrm{frac}_\mathrm{source} &=& 0.44_{-0.15}^{+0.08} ~~, \\
\log_{10}(r_\mathrm{source}) &=& 0.49_{-0.09}^{+0.08} ~~.
\ea

Comparing the case  with $f_{dis} = 1\%$ to  $f_{dis} =10\%$, we clearly see that the fitted parameters are very close. Therefore, in what follows we only consider the case with $f_{dis} = 10\%$.

\section{Two body relaxation using the one dimensional Fokker-Planck approach}
\label{Relaxation}
So far we have fixed the initial conditions of simulation using the actual fits to the observations. Hereafter, we take the median values for the above 6 parameter family and simulate the system for an extended period of time, t = 10.5 \rm{Gyr}s. We aim to study the dynamics of system thoroughly. For this purpose, we consider several quantities including the relaxation timescale, mass growth of the central BH, consumption of the stars and BH to the central BH and the evolution of the density profiles in details. As we will see due to the complexity of having different mass components in the same simulation the evolution is highly non-trivial.

\subsection{Relaxation Timescale}
We start by considering the behavior of the relaxation timescale. 
For a single mass component, it is defined as,

\ba 
\label{Relaxation-Time-Single}
T_{rel}(r)= \frac{\sigma^2}{D[\left(\Delta V_{||} \right)^2]},
\ea
where $\sigma$ refers to the velocity dispersion while $D[\left(\Delta V_{||} \right)^2]$ is given by,
%%%%%%%%%%%%%%%%%%%%%%%%%%%%%%%%%%%%%%%%%%%%%%%%
\begin{figure}
 \center
\includegraphics[width=0.7\textwidth]{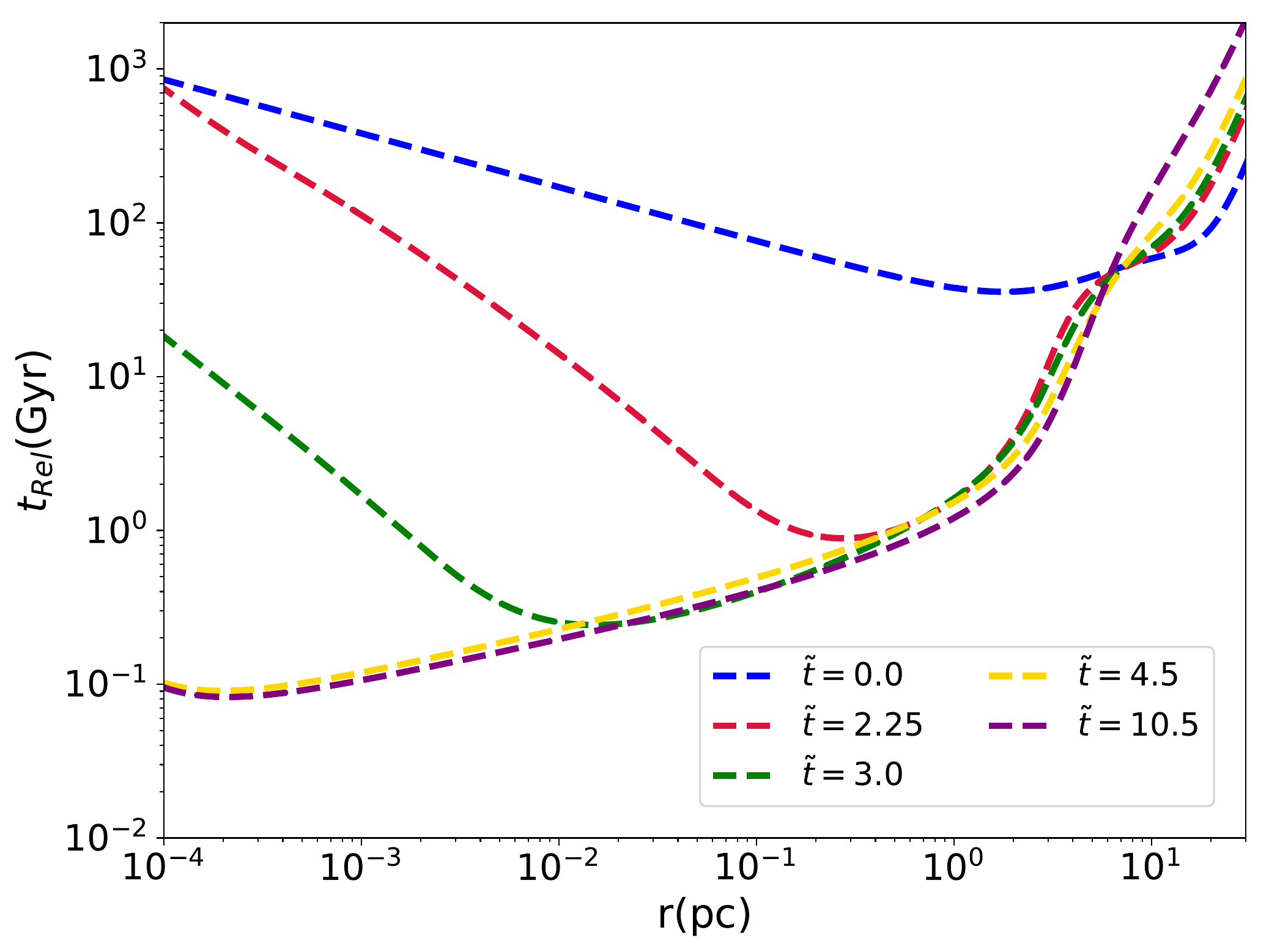}
\caption{Radial dependency of the effective relaxation time-scale at few different times. }
 \label{Relaxation-Time}
\end{figure}
%%%%%%%%%%%%%%%%%%%%%%%%%%%%%%%%%%%%%%%%%%%%%%%%

\ba
\label{velocity-parallel}
D[\left(\Delta V_{||} \right)^2] \simeq 2.94 \frac{G^2 \rho m \ln{\Lambda} }{\sigma},
\ea
In our case, however, we are dealing with multiple masses in the same box and so we should generalize Eq. (\ref{Relaxation-Time-Single}) to case with multiple mass species. There is not a unique way of doing this. Here propose an effective way to compute the relaxation timescale. We compute an averaged velocity dispersion in the numerator of Eq. (\ref{Relaxation-Time-Single}) while we add different contributions in Eq. (\ref{velocity-parallel}). We should however notice that this is only an effective description and in reality heavier BHs may have shorter relaxation timescales. Figure \ref{Relaxation-Time} presents the radial behavior of the effective relaxation timescale for few different times. While the typical relaxation timescale is rather long in the galactic nuclei, it decreases by few orders of magnitude toward smaller radii.

\subsection{Evolution of central BH mass}
\label{evolution-cbh-mass}
\begin{figure}
\includegraphics[width=1.05\textwidth]{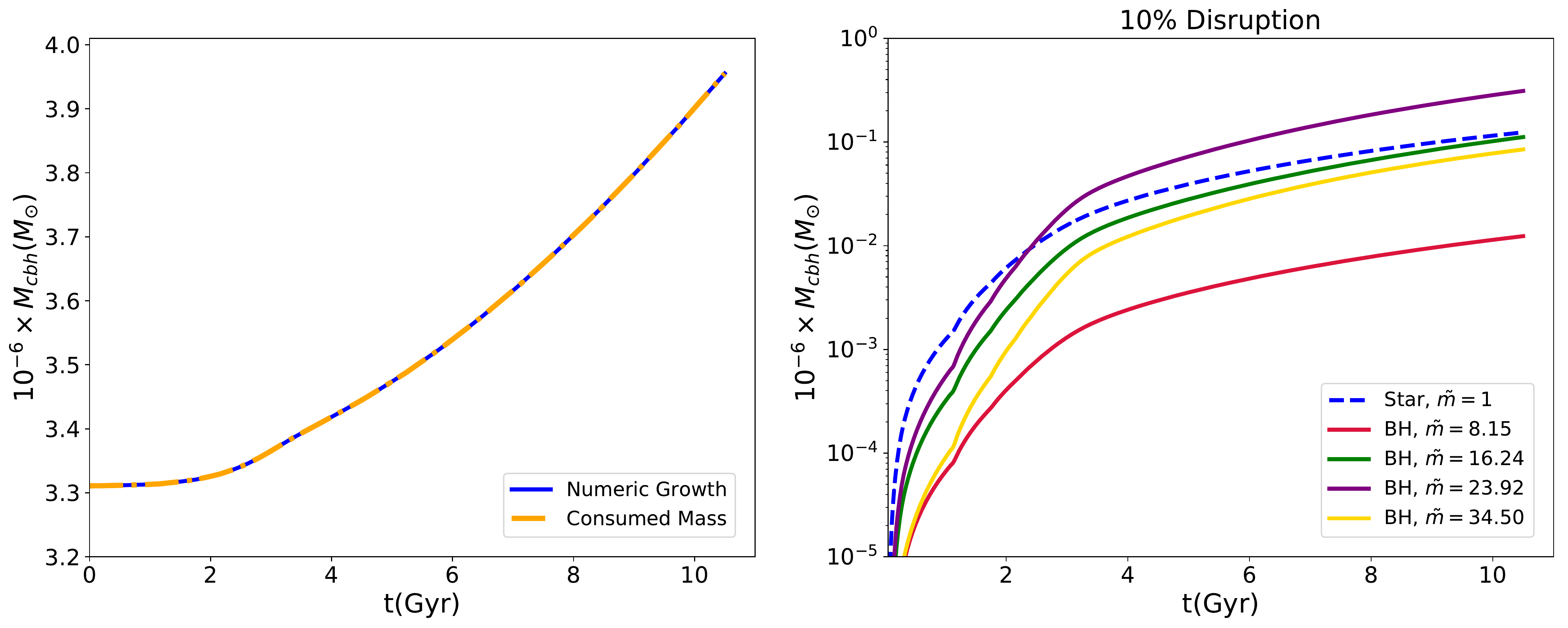}
\caption{Growth of central BH mass with time. On the (left), we show a comparison between the full numeric results with the collection of all of different species. On the (right) panel, we present individual contributions in the mass growth.  }
\label{Mass-Growth}
\end{figure}

Here we study the growth of the mass of central BH with time. Figure \ref{Mass-Growth} presents the dynamical evolution of the central BH mass due to the tidal disruption of stars and BHs with time. Left panel shows a comparison between the full numeric (blue line) with the contribution from different species (both of stars and BHs with different masses) (dotted dashed Orange line). The results are completely matched. According to the plot, the fractional changes in the central BH mass is about 20\% during the entire evolution of the system. 
Right panel shows different components individually. Our results are highly non-trivial. Although we assume a 10\% contribution from the stars, they (almost) dominate the mass growth in the central BH. Although different BHs have the same overall shape, the contribution from the BH with mass $m_{BH} = 23.92 M_{\odot}$, the second heaviest BH, is dominated over the rest of the BHs. A quick look at Tab. \ref{BH-ZAMS-Norm} shows that it is because the initial normalization of this BH, inferred from the BH IMF, is relatively higher than the others. This can be interpreted as the direct impact of the IMF of BH in the evolution of system. We emphasize that this is the first time that the detailed impact of IMF of BH is shown in the tidal disruption event. 

%%%%%%%%%%%%%%%%%%%%%%%%%%%%%%%%%%%%%%%%%%%%%%%%
%%%%%%%%%%%%%%%%%%%%%%%%%%%%%%%%%%%%%%%%%%%%%%%%

\subsection{Averaged mass with radius}
\begin{figure}
 \center
 \includegraphics[width=0.75\textwidth]{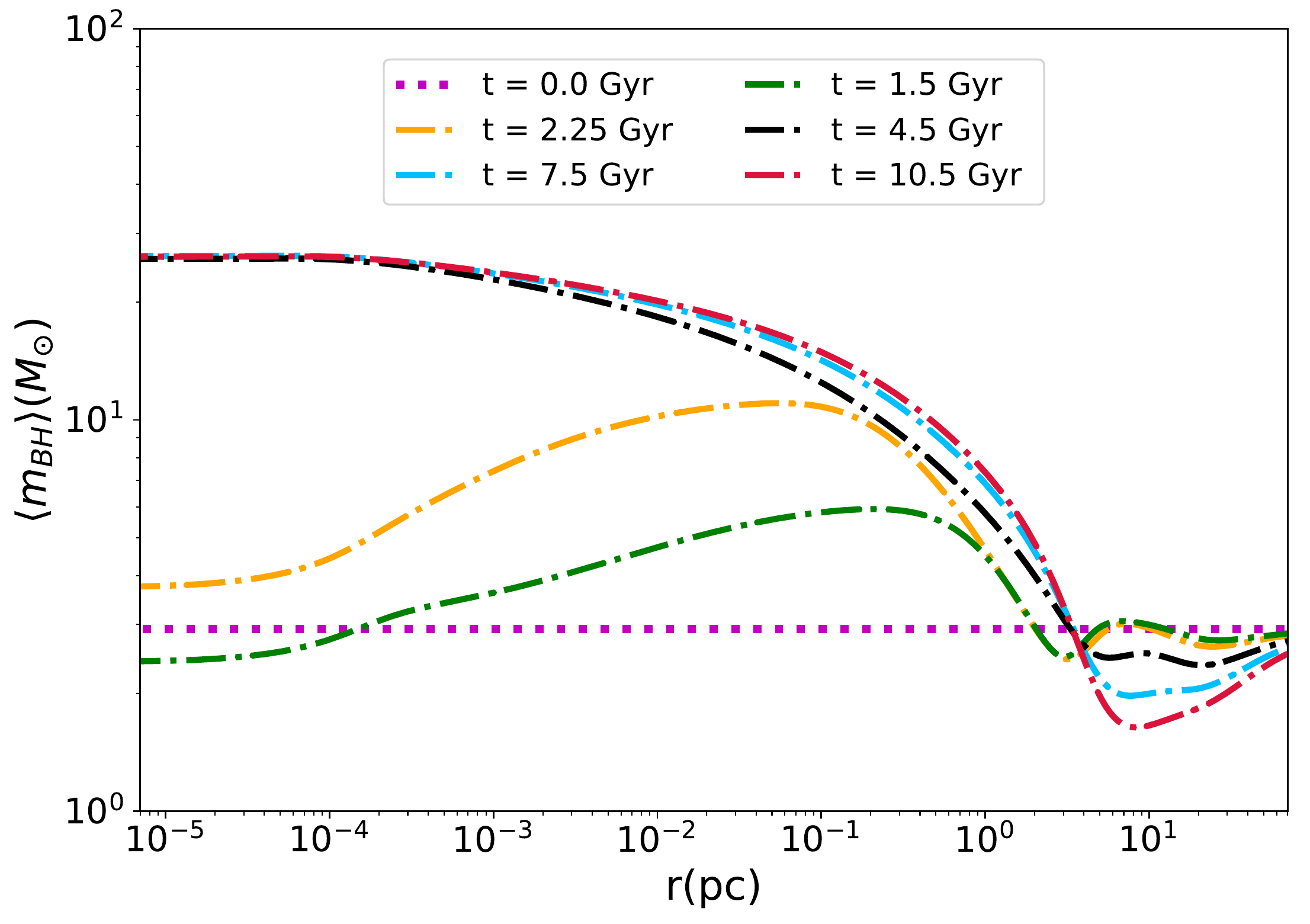}
\caption{Radial profile of the averaged mass around the central BH.}
\label{Mass-Radius-R}
\end{figure}

Next, we consider the radial profile of average masses around the central BH, 
\ba  
\label{mass-radius}
\langle m_{BH}\rangle(r) = \frac{\sum_{i} m_{BH,i} \rho^{i}_{BH}(r)}{\sum_{i} \rho^{i}_{BH}(r)}.
\ea
Figure \ref{Mass-Radius-R} presents the average mass computed using Eq. (\ref{mass-radius}) at different times. From the plot it is clear that averaged mass is quickly redistributed with time early on while shows slower changes later on. Furthermore, while it increases toward the center, it get suppressed at larger radii.

%%%%%%%%%%%%%%%%%%%%%%%%%%%%%%%%%%%%%%%%%%%%%%%%
%%%%%%%%%%%%%%%%%%%%%%%%%%%%%%%%%%%%%%%%%%%%%%%%

\subsection{ 3D/2D density profile of stars and comparison with simulations/observations}

\begin{figure}
 \center
 \includegraphics[width=1.07\textwidth]{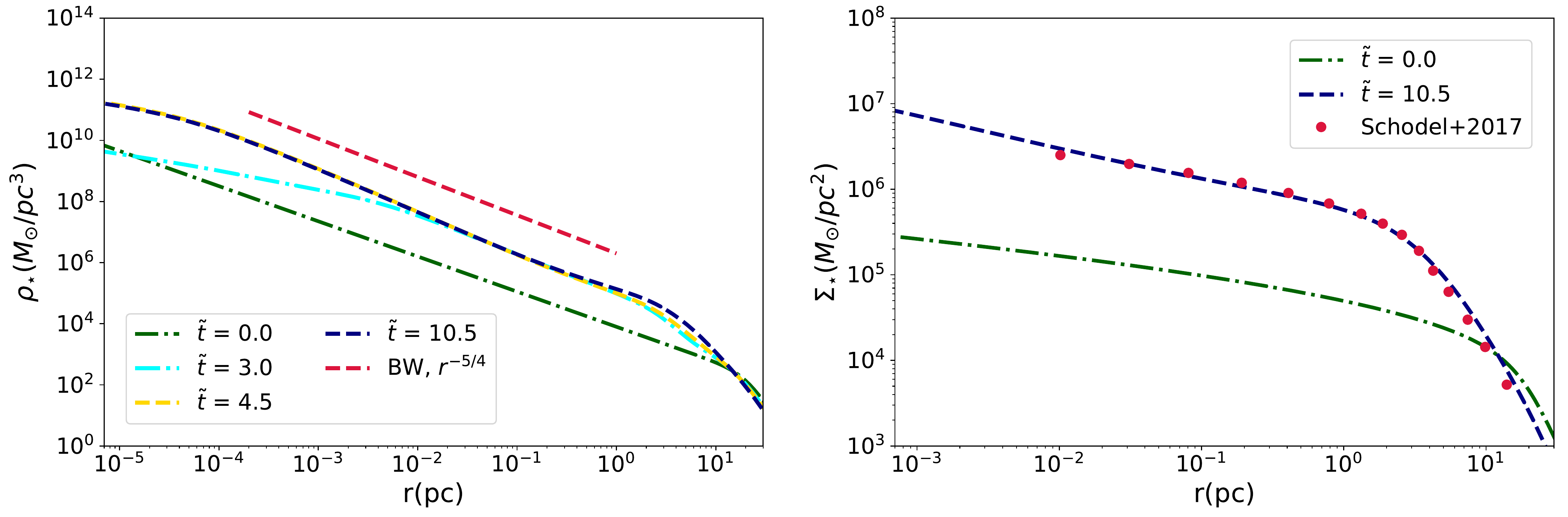}
\caption{(Left) 3 dimensional density profile of stars at different times. Dashed blue line is the fit to the BW cusp with the slope $\gamma = -5/4$. (Right) 2 dimensional surface density profile and comparison of the  final surface density with the recent observational results in \cite{Schodel:2017vjf}.}
\label{2D-3D-star}
\end{figure}

Another interesting quantities are the evolution of the density profile of stars in 2D and 3D. The aim is to compare our results with the recent observational results \cite{Schodel:2017vjf}, in the case of 2D, as well as N-body simulations \cite{Sedda:2018znc, Arca-Sedda:2017wea,Panamarev:2018bwq}. Figure \ref{2D-3D-star} presents the 3D density profile (left) and 2D surface density (right) with time. 
In 3D density profile, dashed red line is a fit to Bahcall-Wolf (BW) cusp \cite{Bahcall-Wolf-1976, Bahcall-Wolf-1977} with the slope $\gamma = -5/4$. 
In 2D surface density profile, we compare the final surface density with the recent observational results in \cite{Schodel:2017vjf}. Our median parameters inferred from MCMC analysis, gives us an incredible fit to the most recent observations in the galactic center.

%%%%%%%%%%%%%%%%%%%%%%%%%%%%%%%%%%%%%%%%%%%%%%%%
%%%%%%%%%%%%%%%%%%%%%%%%%%%%%%%%%%%%%%%%%%%%%%%%

%%%%%%%%%%%%%%%%%%%%%%%%%%%%%%%%%%%%%%%%%%%%%%%
\begin{figure}
 \center
 \includegraphics[width=1.01\textwidth]{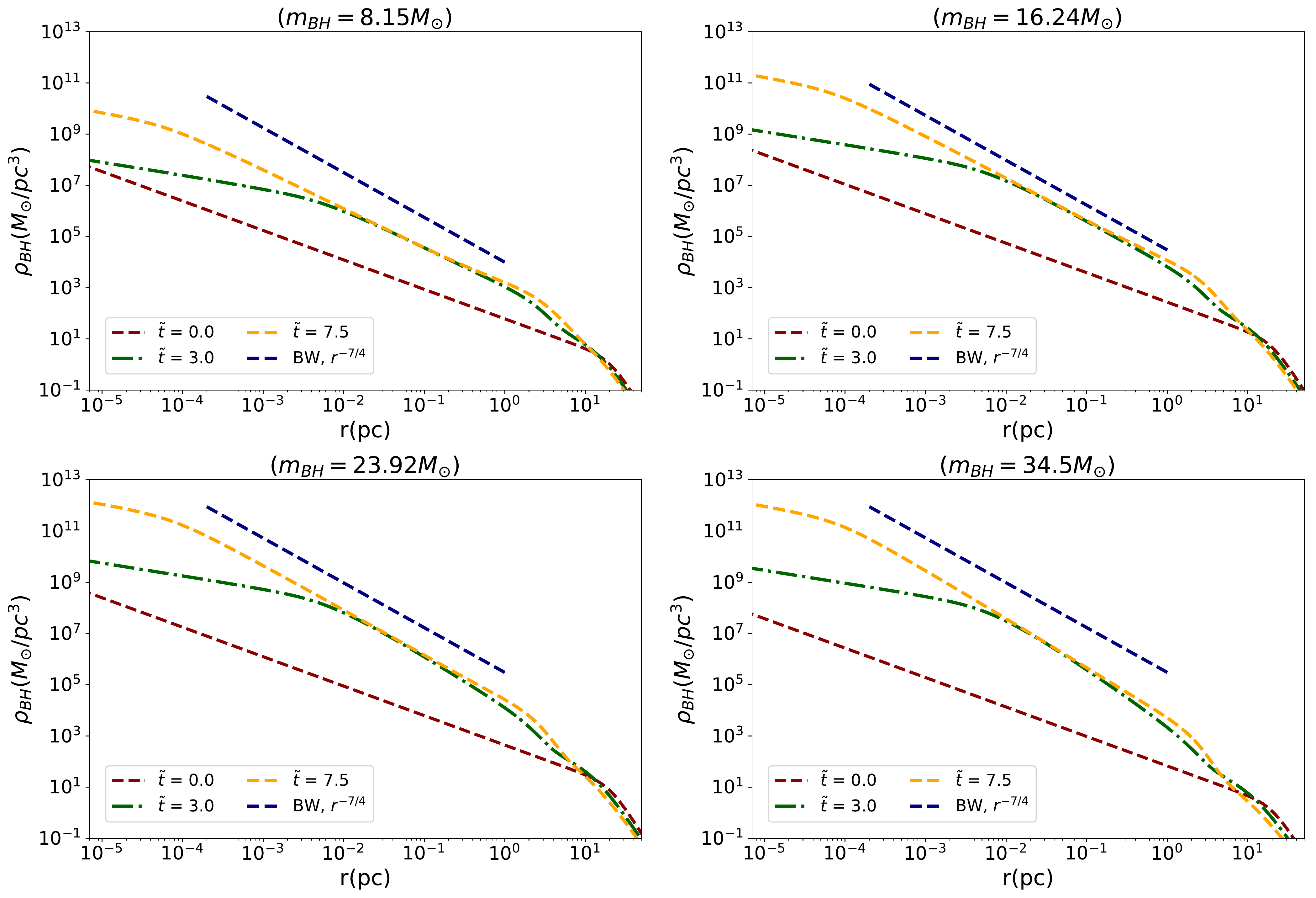}
\caption{Dynamical evolution of the density profile of BHs with different masses with time. }
\label{BH_Density_Profile}
\end{figure}
%%%%%%%%%%%%%%%%%%%%%%%%%%%%%%%%%%%%%%%%%%%%%%%%

 %%%%%%%%%%%%%%%%%%%%%%%%%%%%%%%%%%%%%%%%%%%%%%%%
\begin{figure}
\center
\includegraphics[width=1.01\textwidth]{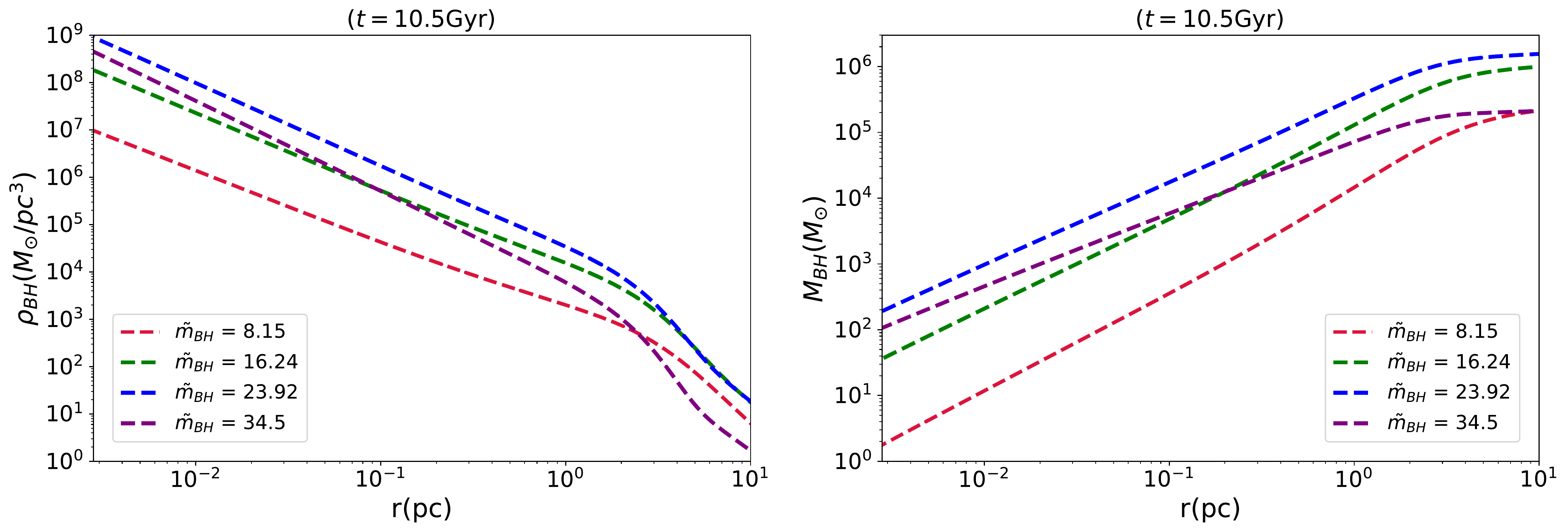}
\caption{Density (left) and mass (right) profile for the BHs after $t = 10.5$ \rm{Gyr}s.}
\label{Final_BH_Density}
\end{figure}
%%%%%%%%%%%%%%%%%%%%%%%%%%%%%%%%%%%%%%%%%%%%%%%%
\subsection{Dynamical evolution of density profile of BHs}
Next we consider the dynamical evolution of BHs density profile with time. Likewise the case of stellar profile, we aim to compare our results with those of N-Body simulations. The main advantage of our approach is that using the Fokker Planck, we may extend the density profiles toward smaller radii which are completely inaccessible with the costly N-Body simulations. Figure \ref{BH_Density_Profile} presents BHs density profiles with time for different BH masses. The growth of the BW cusp is clearly seen from the plot. Since different BH masses are connected through their IMF, we may want to check how does this affect their final density profiles. In figure \ref{Final_BH_Density}, we present density profile of BHs after $t = 10.5$ \rm{Gyr}s.
From the plot, we clearly see the impact of IMF of BHs in enhancing mostly both of the density and mass of second heaviest BH, $m = 23.92 M_{\odot}$ at every radius. Heaviest BH, with $m_{BH} = 34.5 M_{\odot}$ experiences faster movement to the center which leads to further suppression of its density profile at larger radii.

%%%%%%%%%%%%%%%%%%%%%%%%%%%%%%%%%%%%%%%%%%%%%%%%
\begin{figure}
\center
\includegraphics[width=1.01\textwidth]{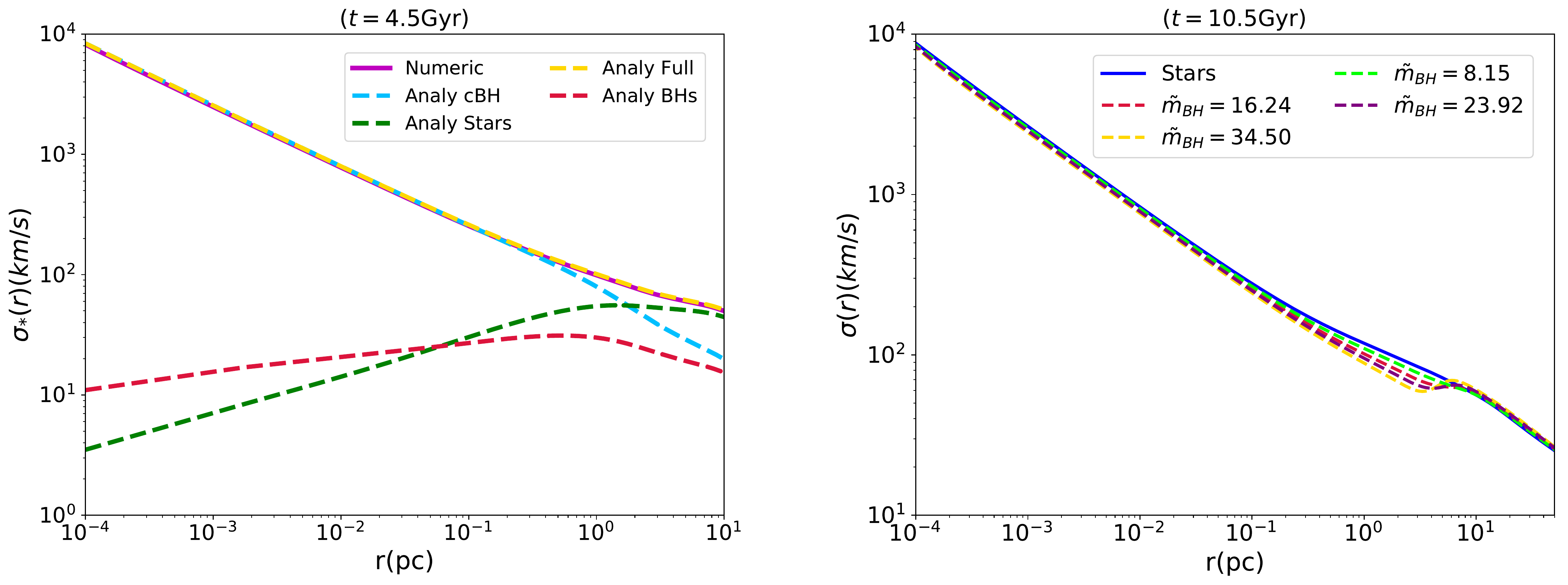}
\caption{Velocity dispersion of the stars and BHs. }
\label{sigma-S-BH}
\end{figure}
%%%%%%%%%%%%%%%%%%%%%%%%%%%%%%%%%%%%%%%%%%%%%%%%

\subsection{Velocity dispersion of BHs and impact of the central BH}
Finally we study the behavior of velocity dispersion defined as,

\ba 
\label{Velocity_Dis}
\sigma_i^2(r) = \frac{G}{\rho_i(r)} \times \int_{r}^{\infty} \frac{\rho_i(r') M(r') }{r'^2} dr', ~~~ i = (star, BH).
\ea
here $\rho_i(r)$ refers to the density profile of stars as well as BHs and $M(r)$ denotes the total mass interior to the radius $r$. As expected there are different contributions in the interior mass including the stars, different BH species and central BH. Figure \ref{sigma-S-BH} presents the radial profile of the velocity dispersion. On the left panel, we present the stellar contribution in the velocity dispersion. From the plot it is clear that at smaller radii, $ r/pc < 0.1 $, the impact of the central BH is dominated in the velocity dispersion. However at larger radii the impact of the stars get dominated thanks to their shallower cusp profile. Right panel compares the velocity dispersion of stars and BHs. It is clearly seen that their velocity profile are quite similar to each other except a little difference for $ 0.1<r/pc < 10$ due to the transition from being dominated by the central BH or stars.

\section{GW Implications}
\label{GW}
So far we presented various aspects of having BH mass-function in the galactic center.
Here we investigate another observational signature of such a mass function in the connection with the GW. We start with the detailed description of the steps involved. Then we present the forecast for the expected number of the events after a maximum operation of LISA of about $T_{obs} = 10 $\rm{yrs}.

\subsection{Set up}
\label{sample}
Here we describe the setting of our simulation for counting the number of expected events as seen with LISA. Our method is based on a post-processing of the AGAMA code where we add in the impact of GW as well as the angular momentum diffusion in the evolution of the semi-major axes, $a$, and eccentricity, $e$. We make a sample of $N_{tot} = 10000$ systems made of pairs of initial $(a, e)$, 100 values in each of them.
Initial values of semi-major axes are taken from a distribution of $f(a)/a$, where $f(a)$ refers to the phase-space distribution function. We allow the system to get relaxed for an amount of $t = 3.7$ \rm{Gyrs} where we see that it does not change dramatically afterward. We use this distribution function in our samples and find a grid of semi-major axes in the range of $(7 \times 10^{-6} - 10)$ pc. In addition, initial values of the eccentricity are chosen from a thermal distribution. Taking these initial conditions, we evolve the system for a maximum period of $t = 6.6 $\rm{Gyrs} or when any given systems cross the loss-cone surface, as is determined in the following.

\subsection{Dynamical Evolution}
As already described above, in our simulations we make a sample of $N_{tot} = 10000$ systems with the above initial conditions. Next, we consider the dynamical evolution of every systems where the dynamics are given as,

\ba 
\label{a-t-relation}
\frac{d}{dt}\left(a(t)\right) &=& -\frac{64}{5} \frac{G^3 m_{bh} M_{\bullet}M_{tot}}{c^5 a^3 (1-e^2)^{7/2}} \left(1 + \frac{73}{24} e^2 + \frac{37}{96} e^4 \right), \\
\label{e-t-relation}
\frac{d}{dt}\left(e^2(t)\right) &=& -\frac{608}{15} \frac{G^3 m_{bh} M_{\bullet}M_{tot} }{c^5 a^4 (1-e^2)^{5/2}} \left(1 + \frac{121}{304} e^2  \right) e^2 + \frac{\mu(a)}{\left(\alpha + \ln{\left(\frac{c^2 a}{16 G M_{\bullet}} \right) \left(1 - e^2 \right) } \right)} .
\ea
where $M_{tot} \equiv m_{bh} + M_{\bullet}$. Here the second term in Eq. (\ref{e-t-relation}) refers to the angular momentum diffusion term , as is discussed in more details at \cite{Emami2020}. We add it for taking into account the impact of the angular momentum diffusion which are averaged in the 1D Fokker Planck approach. Though the details of this term is left to \cite{Emami2020}, 
it is arisen from the loss-rate as described in Eqs. (\ref{loss-rate}) and Eq. (\ref{filling-factor}) with converting the circular orbit, $e =0$, to the elliptical orbit with non-zero eccentricity. It is fair to say that this novel term was ignored in the analysis before and as it turns out this is very important contribution in enhancing the eccentricity and so pushing the system to be potentially detectable with LISA. Though not having the same origin, it is similar to the enhancement in the rate of the GW due to the famous Kozai-Lidov oscillations in the context of the triple systems. Owing to this term, a lot of orbits experience a horizontal evolution, along the eccentricity axes, before they get to the LISA band. This term is responsible in increasing the number of the expected signals in LISA which was almost entirely ignored by the previous authors \cite{AmaroSeoane:2012tx}.

\subsection{Signal to Noise Ratio}
\label{SN-Ratio}

Having presented the simulation set-up as well as the dynamical evolution for the system, here we present our formalism for computing the signal to noise ratio in the LISA band \cite{Fang:2019dnh}

\ba 
\label{SNR1}
\left(SNR\right)^2 = 2 \sum_{n = 1}^{n_{max}} \int \frac{h^2_{c,n}}{ f_n S(f_n)} d \ln{f_n},
\ea
where $n_{max} $ refers to the maximum number of the harmonics. In our analysis we take this number to be maximum $10^5$. In addition, $h_{c,n} = \left( \pi d \right)^{-1} \sqrt{2 \dot{E}_{n}/\dot{f}_n}$. $d$ also refers to the distance from the source. Finally, 
$S(f_n)$ refers to the LISA noise power. 
Since we stop the integration before the loss-cone, we can directly check that in all of our samples we have $\dot{f}/f \ll T_{obs} \simeq 10$yrs. Therefore we can expand the integral in Eq. (\ref{SNR1}) as,

\ba 
\label{SNR2}
\left(SNR\right)^2 = \left(\frac{512}{5 d^2}\right) T_{obs} \frac{\left(G M_{c} \right)^{10/3}}{c^8} \left( 2 \pi f_{orb} \right)^{4/3} \sum_{n = 1}^{n_{max}} \frac{g_n(e)}{n^2 S(f_n)}.
\ea
where $M_c$ refers to the chirp mass and we set $d = 7.8 kpc$. Furthermore, in our estimation, we take the maximum period of LISA operation of about $T_{obs} = 10$ yrs and we choose the criteria of having $SNR = 8 $ in inferring the detectable number of systems.

\subsection{Estimation of rate of events}
\label{Rate-Estimation}
Here we combine the above individual pieces and describe our method in estimating the total rate of expected LISA events after $T = 10$ yrs of LISA operations, where we have taken the maximum duration of LISA to increase the chance of detecting LISA sources. 

Starting from the initial condition for every pairs of $(a,e)$ as mentioned above, we evolve every systems with time either to the time that they cross the loss-cone area, defined as $a(1-e) = 8 G M_{\bullet}$, or up to $t = 6.6$ \rm{Gyrs}. Using Eq. (\ref{SNR2}), we compute the signal to noise ratio for each of the above systems during the entire of their lifetime, defined from $t = 0 $ to either their loss-cone crossing time or up to $t = 6.6$ \rm{Gyrs} depending on which happens earlier. For every systems, we compute the total duration of time that the system has $SNR \geq 8$ with a frequency in the LISA band. We call this the lifetime of being in the LISA and 
divide it to the total lifetime of this system. For example, if the system has a total lifetime of $t = 10^{-6}$ \rm{Gyrs} and spends a period of $\Delta t = 10^{-7}$ \rm{Gyrs} in the LISA band, the weighting factor for this system would be $w = 10^{-7}/10^{-6}$.
Next, we should take into account the rate of the replenishment for every systems. Since the replenishment is due to the energy diffusion, we use the diffusion rate as an extra factor for every systems. Therefore we get a combined rate factor, hereafter $w(i)$, for ith system as,

\ba 
\label{Total-Weight}
W(i) = \left(\frac{\Delta t_i}{t_i} \right) \times \mu_i,
\ea
where $\Delta t_i$, $t_i$ and $\mu_i$ refer to the spent period of time in the LISA band, the total lifetime and the diffusion rate for i-th system, respectively. 

At the next level, we should rescale the above number to the actual number of the BHs in the galactic center interior to the radius $r = r_{sam} = 10 pc$, the upper limit in our sample making,  as describe in \ref{sample}. This can be done by the factor $N_{BH}(r \leq r_{sam})/N_{tot}$ where $N_{BH}(r \leq r_{sam}) \equiv M_{BH}(r \leq r_{sam})/m_{BH}$ refers to the number of the BHs with mass $m_{BH}$ interior to the radius $r = r_{sam} = 10 pc$ and $N_{tot} = 10000$ is the total number of the samples in our system. Therefore, the final rescaled rate for i-th system would be, hereafter $W_{tot}(i)$,

\ba 
\label{Total-Weight}
W_{tot}(i) = \left(\frac{N_{BH}(r \leq r_{sam})}{N_{tot}} \right) \times W(i),
\ea

Furthermore, we should also eliminate the cases with semi-major axes above a critical value, hereafter $a_{crit}$, where the timescale of the GW, $\tau_{GW} \equiv |a/\dot{a}|$, is longer than the loss-cone timescale $t_{LC} \equiv \left(L_{LC}/L_{cir} \right)^2 t_{Rel}$. Where the loss-cone timescale is defined as a timescale associated with a change in the angular momentum by the order of the loss-cone \cite{Hopman:2006xn}. 
Combining with the loss-cone surface, $a(1-e) = 8 G M_{\bullet}$, we can estimate the critical semi-major axes above which $\tau_{GW} > t_{LC}$. As argued in \cite{Hopman:2006xn}, systems with $a>a_{crit}$ scatter either inward the loss-cone or to larger orbits without emitting the GWs. Therefore they are not source of the inspiral phase and must be removed from our sample. In Table \ref{a-crit_tab11} we estimate $a_{crit}$ for different BH masses.

%%%%%%%%%%%%%%%%%%%%%%%%%%%%%%%%%%%%%%%%%%%%%%%%%%
\begin{table}
\centering
\caption{Critical value of semi-major axes above which we do not have any inspiral phase.}
\label{a-crit_tab11}
\begin{tabular}{lccr} 
		\hline
		\hline
		 $m_{BH} (M_{\odot})$ & 
        $a_{crit}(pc)$ &
        \\
		\hline
       $8.15$ & 
        $0.01$ &
        \\
        \hline
        $16.24$ & 
        $~ 0.029$ & 
        \\
        \hline
        $23.92$ & 
        $0.04$ &
        \\
        \hline
        $34.50$ & 
        $0.05$ 
        \\
        \hline
    	\end{tabular}
\end{table}
%%%%%%%%%%%%%%%%%%%%%%%%%%%%%%%%%%%%%%%%%%%%%%%%%%%%

In order to get the rate of the observable systems in our sample, $R_{obs}$, we sum over the total rescaled rate of any systems as, 

\ba 
\label{Observable-Fraction}
R_{obs} &\equiv& \sum_i W_{tot}(i) \nonumber\\
&=& \sum_i  \left(\frac{N_{BH}(r \leq r_{sam})}{N_{tot}} \right)  \left(\frac{\Delta t_i}{t_i} \right) \mu_i, 
\ea

The total number of the LISA inspiral events after $T_{obs} = 10 $ \rm{yrs} would be,

\ba 
\label{tot-Number1}
N_{lisa} = R_{obs} \times (10 \rm{yrs}).
\ea

\subsection{Results and Discussions}
\label{Results}
Having described our novel formalism in computing the total expected numbers of the LISA events, here we present the final number for our system of 4 BHs. Table \ref{number} presents the expected inspiral rate with LISA as well as the total number of LISA sources from the inspiralling BHs from the galactic center after $T_{obs} = 10 $ \rm{yrs}. There are few important points that are worth highlighting here, 

First, the rate for one galaxy is rather small. Therefore, it is of interest to use this novel method for different galaxies, taking into account their internal structures, and read off the final expected rate and number. If we focus on Milky Way like galaxies, with similar black holes to SgrA*, we could imagine that combining $N_{gal} \sim 10^4$ of them we may be able to see one inspiralling signal. Care should be taken for other type of galaxies with different structures, though. In \cite{Emami2020}, we generalize this formalism to more galaxies as well as larger samples of the BHs in hand and also different metallicities. 

 Second, the impact of the BH mass-function is clearly seen in the total expected number of events in LISA. Indeed, comparing Table \ref{BH-ZAMS-Norm} with Table \ref{number}, we clearly see that the total rate is peaked at the peak of the mass-function. This gives us a another signature of the BH mass-function this time in the GW signal. 

%%%%%%%%%%%%%%%%%%%%%%%%%%%%%%%%%%%%%%%%%%%%%%%%%%
\begin{table}
\centering
\caption{Expected rate of the inspiral as well as the number of LISA sources per one MW like galaxy.}
\label{number}
\begin{tabular}{lccr} 
		\hline
		\hline
		 $ m_{BH} (M_{\odot})$ & 
        $ R_{obs} (yr^{-1})$ &
        $ N_{lisa} (T_{obs} = 10 yr)$
        \\
		\hline
       $8.15$ & 
        $2.0 \times 10^{-6}$ &
        $2.0 \times 10^{-5}$ 
        \\
        \hline
        $16.24$ & 
        $1.54 \times 10^{-5} $ & 
        $1.54 \times 10^{-4} $
        \\
        \hline
        $23.92$ & 
        $2.9 \times 10^{-5}$ &
        $2.9 \times 10^{-4} $
        \\
        \hline
        $34.50$ & 
        $5.6 \times 10^{-6}$ &
         $5.6 \times 10^{-5}$
        \\
        \hline
    	\end{tabular}
\end{table}
%%%%%%%%%%%%%%%%%%%%%%%%%%%%%%%%%%%%%%%%%%%%%%%%%%%%

\section{Conclusions}
\label{conclusion}
Using an orbit averaged tool, AGAMA code, we simulated the galactic center in the presence of stars in the main sequence and 4 different BH species. We used a binary stellar evolution code, COSMIC, to create a sample of BHs from an initial low metallicity of $Z = 0.001$ and constructed the BH mass-function initially. For the first time in studies of  galactic nuclei, we explored the impact of the BH mass-function in the evolution of the system. We have also added the impact of the continuous star formations in the system. 
We used an MCMC approach for computing the initial profile of stars and BHs which are matched with the current observations after a total duration of $t = 10$ \rm{Gyrs}. We implemented the above results in the Phase-Flow code, a library in the AGAMA code, and studied the dynamical evolution of this system precisely. The novelty of this work involves the following results.

$\bullet$ First , the above hybrid tool of the stellar evolution with the AGAMA code enabled us to figure out the impact of the BH mass-function in the evolution of the system through its direct effect on the initial normalization of different BH species. As shown in Table \ref{BH-ZAMS-Norm}, the normalization of the second heaviest BH is dominated over the rest of the BHs thanks to a local peak in the BH mass-function. 

$\bullet$ Second, we explicitly showed that it leads to an enhancement in the contribution of the second heaviest BH in the growth of the central BH's mass as they are more abundant in the system. Quite interestingly, as shown in the right panel of Figure \ref{Mass-Growth}, taking  $f_{dis} = 10\%$, stars make the second important contribution in the mass of the central BH. 

$\bullet$ Third, detailed study of the velocity dispersion profile revealed that very close to the Central BH, stars and BHs behave similarly in the velocity dispersion. This would change in some intermediate scales where the contribution of stars begin to dominate over the contribution of the central BH which leads to a deviation from the Keplerian orbits thanks to shallower cuspy profile of stars compared with the BHs. Observationally it is very interesting and may enable us to measure the cusp profile by looking at the projected velocity dispersion close in the center. We leave further investigation of this to a future work.  
getting important. Finally, at larger radii stars are the dominant source of the velocity dispersion. 

$\bullet$ Finally, we considered the GW in this system. Since the orbit averaged Phase Flow does not consider the angular momentum diffusion nore the GW dissipation, we made a post-processing analysis where we took into account both of these effects self consistently. For the first time in this context, we considered these effects collectively and explicitly showed an enhancement in the eccentricity from the angular momentum diffusion. We made a sample of 10000 pairs in the initial semi-major axes and eccentricity and evolved each of them for a maximum period of $t = 6.6$ \rm{Gyrs}. We computed the SNR for each of them during the entire of their lifetime and computed the fraction of the time that they spend in the LISA band, defined with the criteria of having $SNR \geq 8$ and in the LISA frequency band to their entire lifetime. WE weighted this number with the diffusion rate as the replenishment factor. We also rescaled that number with the actual number of the BHs interior to $r = 10 pc$ as the upper limit in our sample making. We presented the total rate at Eq. (\ref{Observable-Fraction}). 
Furthermore, we also eliminated the systems with $a>a_{crit}$ where $a_{crit}$ refers to the maximum semi-major axes that above it the characteristic timescale of the GW is above the loss-cone time-scale. BHs in this area either scatter off the central BH to larger orbits or get swallowed to the center without emitting any GWs. We removed these systems from our samples entirely.
Our computation showed that after a maximum duration of $T_{obs} = 10$ yrs, we get a rate of $10^{-6}-10^{-5}$ 1/yrs as the inspiralling rate for the BHs with different masses. 
The total expected rate and number of one MW like galaxy is given in 
Table \ref{number}. This means that in order to get one signal from the inspiral phase we need to consider of order of $N_{gal} \simeq 10^4  $ MW like galaxies. Quite interestingly the inspiral rate is peaked at the peak of the BH mass-function. This is the second signature of the the BH mass-function proposed here.

\section*{Acknowledgment}
We thank Christopher Berry, Sownak Bose, John Forbes, Dan D'Orazio and Morgan MacLeod for helpful discussions.
We especially thanks Eugene Vasiliev and Fabio Antonini for very insightful discussions and for a followup collaboration. We are grateful to the anonymous referee for their very constructive comments. 
R.E. acknowledges the support by the Institute for Theory and Computation at Center for Astrophysics, Harvard-Smithsonian. This work was also supported in part by the Black Hole Initiative at Harvard University which is funded by a JTF grant. 

\newpage

%%%%%%%%%%%%%%%%%%%%%%%%%%%%%%%%%%%%%%%%%%%%%%%%


\begin{thebibliography}{!}

%\cite{Bahcall-Wolf-1976}
 \bibitem{Bahcall-Wolf-1976}
 % ``STAR DISTRIBUTION AROUND A MASSIVE BLACK HOLE IN A GLOBULAR CLUSTER"
Bahcall J. N., Wolf R. A., 1976, ApJ, 209, 214

%\cite{Freitag:2006qf}
\bibitem{Freitag:2006qf} 
  M.~Freitag, P.~Amaro-Seoane and V.~Kalogera,
  %``Stellar remnants in galactic nuclei: mass segregation,''
  Astrophys.\ J.\  {\bf 649}, 91 (2006)
  doi:10.1086/506193
  [astro-ph/0603280].
  %%CITATION = doi:10.1086/506193;%%
  %139 citations counted in INSPIRE as of 17 Feb 2019


%\cite{Alexander:2008tq}
\bibitem{Alexander:2008tq} 
  T.~Alexander and C.~Hopman,
  %``Strong mass segregation around a massive black hole,''
  Astrophys.\ J.\  {\bf 697}, 1861 (2009)
  doi:10.1088/0004-637X/697/2/1861
  [arXiv:0808.3150 [astro-ph]].
  
  %\cite{Hopman:2006xn}
\bibitem{Hopman:2006xn} 
  C.~Hopman and T.~Alexander,
  %``The effect of mass-segregation on gravitational wave sources near massive black holes,''
  Astrophys.\ J.\  {\bf 645}, L133 (2006)
  doi:10.1086/506273
  [astro-ph/0603324].


  
  %\cite{Hopman:2005vr}
\bibitem{Hopman:2005vr} 
  C.~Hopman and T.~Alexander,
  %``The Orbital statistics of stellar inspiral and relaxation near a massive black hole: Characterizing gravitational wave sources,''
  Astrophys.\ J.\  {\bf 629}, 362 (2005)
  doi:10.1086/431475
  [astro-ph/0503672].
  %\cite{Preto:2009kd}
\bibitem{Preto:2009kd} 
  M.~Preto and P.~Amaro-Seoane,
  %``On strong mass segregation around a massive black hole: Implications for lower-frequency gravitational-wave astrophysics,''
  Astrophys.\ J.\  {\bf 708}, L42 (2010)
  doi:10.1088/2041-8205/708/1/L42
  [arXiv:0910.3206 [astro-ph.GA]].
%\cite{Alexander:2017arl}
\bibitem{Alexander:2017arl} 
  T.~Alexander and B.~Bar-Or,
  %``A universal minimal mass scale for present-day central black holes,''
  arXiv:1701.00415 [astro-ph.GA].
  %%CITATION = ARXIV:1701.00415;%%  
 
 %\cite{Babak:2017tow}
\bibitem{Babak:2017tow} 
  S.~Babak {\it et al.},
  %``Science with the space-based interferometer LISA. V: Extreme mass-ratio inspirals,''
  Phys.\ Rev.\ D {\bf 95}, no. 10, 103012 (2017)
  doi:10.1103/PhysRevD.95.103012
  [arXiv:1703.09722 [gr-qc]].


%\cite{Aharon:2016kil}
\bibitem{Aharon:2016kil} 
  D.~Aharon and H.~B.~Perets,
  %``The impact of mass segregation and star-formation on the rates of gravitational-wave sources from extreme mass ratio inspirals,''
  Astrophys.\ J.\  {\bf 830}, no. 1, L1 (2016)
  doi:10.3847/2041-8205/830/1/L1
  [arXiv:1609.01715 [astro-ph.GA]].


%\cite{OLeary:2008myb}
\bibitem{OLeary:2008myb} 
  R.~M.~O'Leary, B.~Kocsis and A.~Loeb,
  %``Gravitational waves from scattering of stellar-mass black holes in galactic nuclei,''
  Mon.\ Not.\ Roy.\ Astron.\ Soc.\  {\bf 395}, no. 4, 2127 (2009)
  doi:10.1111/j.1365-2966.2009.14653.x
  [arXiv:0807.2638 [astro-ph]].


%\cite{Banerjee:2009hs}
\bibitem{Banerjee:2009hs} 
  S.~Banerjee, H.~Baumgardt and P.~Kroupa,
  %``Stellar-mass black holes in star clusters: implications for gravitational wave radiation,''
  Mon.\ Not.\ Roy.\ Astron.\ Soc.\  {\bf 402}, 371 (2010)
  doi:10.1111/j.1365-2966.2009.15880.x
  [arXiv:0910.3954 [astro-ph.SR]].


%\cite{Antonini:2012ad}
\bibitem{Antonini:2012ad} 
  F.~Antonini and H.~B.~Perets,
  %``Secular evolution of compact binaries near massive black holes: Gravitational wave sources and other exotica,''
  Astrophys.\ J.\  {\bf 757}, 27 (2012)
  doi:10.1088/0004-637X/757/1/27
  [arXiv:1203.2938 [astro-ph.GA]].


%\cite{Boehle}
\bibitem{Boehle} 
A.~ Boehle, A.~M.~Ghez, R.~Schödel, L.~Meyer, S.~Yelda, S.~Albers, G.~D.~Martinez, E.~E.~Becklin, T.~Do, J.~R.
~Lu, K.~Matthews, M.~R.~Morris, B.~Sitarski and G.~Witzel,
%An Improved Distance and Mass Estimate for Sgr A* from a Multistar Orbit Analysis,
Astrophysical Journal/830/1

%\cite{Berry:2013ara}
\bibitem{Berry:2013ara} 
  C.~P.~L.~Berry and J.~R.~Gair,
  %``Expectations for extreme-mass-ratio bursts from the Galactic Centre,''
  Mon.\ Not.\ Roy.\ Astron.\ Soc.\  {\bf 435}, 3521 (2013)
  doi:10.1093/mnras/stt1543
  [arXiv:1307.7276 [astro-ph.HE]].

%\cite{Antonini:2016gqe}
\bibitem{Antonini:2016gqe} 
  F.~Antonini and F.~A.~Rasio,
  %``Merging black hole binaries in galactic nuclei: implications for advanced-LIGO detections,''
  Astrophys.\ J.\  {\bf 831}, no. 2, 187 (2016)
  doi:10.3847/0004-637X/831/2/187
  [arXiv:1606.04889 [astro-ph.HE]].

%\cite{Generozov:2018niv}
\bibitem{Generozov:2018niv} 
  A.~Generozov, N.~C.~Stone, B.~D.~Metzger and J.~P.~Ostriker,
  %``An overabundance of black hole X-ray binaries in the Galactic Centre from tidal captures,''
  Mon.\ Not.\ Roy.\ Astron.\ Soc.\  {\bf 478}, no. 3, 4030 (2018)
  doi:10.1093/mnras/sty1262
  [arXiv:1804.01543 [astro-ph.HE]].



%\cite{Auchettl:2016qfa}
\bibitem{Auchettl:2016qfa} 
  K.~Auchettl, J.~Guillochon and E.~Ramirez-Ruiz,
  %``New physical insights about Tidal Disruption Events from a comprehensive observational inventory at X-ray wavelengths,''
  Astrophys.\ J.\  {\bf 838}, no. 2, 149 (2017)
  doi:10.3847/1538-4357/aa633b
  [arXiv:1611.02291 [astro-ph.HE]].
  %%CITATION = doi:10.3847/1538-4357/aa633b;%%
  %44 citations counted in INSPIRE as of 17 Feb 2019

%\cite{Arca-Sedda:2017qcq}
\bibitem{Arca-Sedda:2017qcq} 
  M.~Arca-Sedda, B.~Kocsis and T.~Brandt,
  %``Gamma-ray and X-ray emission from the Galactic centre: hints on the nuclear star cluster formation history,''
  Mon.\ Not.\ Roy.\ Astron.\ Soc.\  {\bf 479}, no. 1, 900 (2018)
  doi:10.1093/mnras/sty1454
  [arXiv:1709.03119 [astro-ph.GA]].
  %%CITATION = doi:10.1093/mnras/sty1454;%%
  %15 citations counted in INSPIRE as of 17 Feb 2019

%\cite{Perna:2019axr}
\bibitem{Perna:2019axr} 
  R.~Perna, Y.~H.~Wang, N.~Leigh and M.~Cantiello,
  %``On the Apparent Dichotomy Between the Masses of Black Holes Inferred via X-rays and via Gravitational Waves,''
  arXiv:1901.03345 [astro-ph.HE].

%\cite{Bortolas:2017moe}
\bibitem{Bortolas:2017moe} 
  E.~Bortolas, M.~Mapelli and M.~Spera,
  %``Supernova Kicks and Dynamics of Compact Remnants in the Galactic Centre,''
  Mon.\ Not.\ Roy.\ Astron.\ Soc.\  {\bf 469}, no. 2, 1510 (2017)
  doi:10.1093/mnras/stx930
  [arXiv:1704.05850 [astro-ph.GA]].  

%\cite{Fragione:2018yrb}
\bibitem{Fragione:2018yrb} 
  G.~Fragione, E.~Grishin, N.~W.~C.~Leigh, H.~B.~Perets and R.~Perna,
  %``Black Hole and Neutron Star Mergers in Galactic Nuclei,''
  arXiv:1811.10627 [astro-ph.GA].

%\cite{Chatterjee:2001ty}
\bibitem{Chatterjee:2001ty} 
  P.~Chatterjee, L.~Hernquist and A.~Loeb,
  %``Dynamics of a massive black hole at the center of a dense stellar system,''
  Astrophys.\ J.\  {\bf 572}, 371 (2002)
  doi:10.1086/340224
  [astro-ph/0107287].

%\cite{Chatterjee:2002bg}
\bibitem{Chatterjee:2002bg} 
  P.~Chatterjee, L.~Hernquist and A.~Loeb,
  %``Brownian motion in gravitationally-interacting systems,''
  Phys.\ Rev.\ Lett.\  {\bf 88}, 121103 (2002)
  doi:10.1103/PhysRevLett.88.121103
  [astro-ph/0202257].

%\cite{Vasiliev:2017sbo}
\bibitem{Vasiliev:2017sbo} 
  E.~Vasiliev,
  %``A New Fokker–Planck Approach for the Relaxation-driven Evolution of Galactic Nuclei,''
  Astrophys.\ J.\  {\bf 848}, no. 1, 10 (2017)
  doi:10.3847/1538-4357/aa8cc8
  [arXiv:1709.04467 [astro-ph.GA]].
  %%CITATION = doi:10.3847/1538-4357/aa8cc8;%%
  %5 citations counted in INSPIRE as of 19 Dec 2018

% \cite{Binney-Tremaine}
 \bibitem{Binney-Tremaine}
Galactic Dynamics: 
Second Edition, by James Binney and Scott Tremaine. ISBN 978-0-691-13026-2 (HB). Published by Princeton University Press, Princeton, NJ USA, 2008.

%\cite{Rubin:2010bw}
\bibitem{Rubin:2010bw} 
  D.~Rubin and A.~Loeb,
  %``Constraining the Stellar Mass Function in the Galactic Center via Mass Loss from Stellar Collisions,''
  Adv.\ Astron.\  {\bf 2011}, 174105 (2011)
  doi:10.1155/2011/174105
  [arXiv:1012.0583 [astro-ph.GA]].	

%\cite{Belczynski:2016obo}
\bibitem{Belczynski:2016obo} 
  K.~Belczynski, D.~E.~Holz, T.~Bulik and R.~O'Shaughnessy,
  %``The first gravitational-wave source from the isolated evolution of two 40-100 Msun stars,''
  Nature {\bf 534}, 512 (2016)
  doi:10.1038/nature18322
  [arXiv:1602.04531 [astro-ph.HE]].

%\cite{Bortolas:2019sif}
\bibitem{Bortolas:2019sif} 
  E.~Bortolas and M.~Mapelli,
  %``Can supernova kicks trigger EMRIs in the Galactic Centre?,''
  Mon.\ Not.\ Roy.\ Astron.\ Soc.\  {\bf 485}, no. 2, 2125 (2019)
  doi:10.1093/mnras/stz440
  [arXiv:1902.04581 [astro-ph.GA]].


%\cite{Levin:2003kp}
\bibitem{Levin:2003kp} 
  Y.~Levin and A.~M.~Beloborodov,
  %``Stellar disk in the galactic center - A Remnant of a dense accretion disk?,''
  Astrophys.\ J.\  {\bf 590}, L33 (2003)
  doi:10.1086/376675
  [astro-ph/0303436].

%\cite{Aharon:2014tpa}
\bibitem{Aharon:2014tpa} 
  D.~Aharon and H.~B.~Perets,
  %``Formation and evolution of nuclear star clusters with in-situ star-formation: Nuclear cores and age segregation,''
  Astrophys.\ J.\  {\bf 799}, no. 2, 185 (2015)
  doi:10.1088/0004-637X/799/2/185
  [arXiv:1409.5121 [astro-ph.GA]].

%\cite{Kennedy:2016tyr}
\bibitem{Kennedy:2016tyr} 
  G.~F.~Kennedy, Y.~Meiron, B.~Shukirgaliyev, T.~Panamarev, P.~Berczik, A.~Just and R.~Spurzem,
  %``Star–disc interaction in galactic nuclei: orbits and rates of accreted stars,''
  Mon.\ Not.\ Roy.\ Astron.\ Soc.\  {\bf 460}, no. 1, 240 (2016)
  doi:10.1093/mnras/stw908
  [arXiv:1604.05309 [astro-ph.GA]].

%\cite{Karas:2007ds}
\bibitem{Karas:2007ds} 
  V.~Karas and L.~Subr,
  %``Enhanced activity of massive black holes by stellar capture assisted by a self-gravitating accretion disc,''
  Astron.\ Astrophys.\  {\bf 470}, 11 (2007)
  doi:10.1051/0004-6361:20066068
  [arXiv:0704.2781 [astro-ph]].

%\cite{Hopman:2006qr}
\bibitem{Hopman:2006qr} 
  C.~Hopman and T.~Alexander,
  %``Resonant relaxation near a massive black hole: the stellar distribution and gravitational wave sources,''
  Astrophys.\ J.\  {\bf 645}, 1152 (2006)
  doi:10.1086/504400
  [astro-ph/0601161].


%\cite{Merritt:2011ve}
\bibitem{Merritt:2011ve} 
  D.~Merritt, T.~Alexander, S.~Mikkola and C.~M.~Will,
  %``Stellar Dynamics of Extreme-Mass-Ratio Inspirals,''
  Phys.\ Rev.\ D {\bf 84}, 044024 (2011)
  doi:10.1103/PhysRevD.84.044024
  [arXiv:1102.3180 [astro-ph.CO]].

%\cite{Bar-Or:2016qop}
\bibitem{Bar-Or:2016qop} 
  B.~Bar-Or and T.~Alexander,
  %``Steady-state Relativistic Stellar Dynamics Around a Massive Black Hole,''
  Astrophys.\ J.\  {\bf 820}, no. 2, 129 (2016)
  doi:10.3847/0004-637X/820/2/129
  [arXiv:1508.01390 [astro-ph.GA]].

%\cite{Szolgyen:2018zra}
\bibitem{Szolgyen:2018zra} 
  Á.~Szölgyén and B.~Kocsis,
  %``Black Hole Disks in Galactic Nuclei,''
  Phys.\ Rev.\ Lett.\  {\bf 121}, no. 10, 101101 (2018)
  doi:10.1103/PhysRevLett.121.101101
  [arXiv:1803.07090 [astro-ph.GA]].


%\cite{Merritt:2015vxa}
\bibitem{Merritt:2015vxa} 
  D.~Merritt,
  %``Gravitational Encounters and the Evolution of Galactic Nuclei. I. Method,''
  Astrophys.\ J.\  {\bf 804}, 52 (2015)
  doi:10.1088/0004-637X/804/1/52
  [arXiv:1505.07516 [astro-ph.GA]].

%\cite{Merritt:2015xpa}
\bibitem{Merritt:2015xpa} 
  D.~Merritt,
  %``Gravitational Encounters and the Evolution of Galactic Nuclei. II. Classical and Resonant Relaxation,''
  Astrophys.\ J.\  {\bf 804}, 128 (2015)
  doi:10.1088/0004-637X/804/2/128
  [arXiv:1506.03010 [astro-ph.GA]].
  
  
  %\cite{Merritt:2015elb}
\bibitem{Merritt:2015elb} 
  D.~Merritt,
  %``Gravitational Encounters and the Evolution of Galactic Nuclei. III. Anomalous Relaxation,''
  Astrophys.\ J.\  {\bf 810}, 2 (2015)
  doi:10.1088/0004-637X/810/1/2
  [arXiv:1509.01263 [astro-ph.GA]].

%\cite{Merritt:2015kba}
\bibitem{Merritt:2015kba} 
  D.~Merritt,
  %``Gravitational Encounters and the Evolution of Galactic Nuclei. iv. Captures Mediated by Gravitational-wave Energy Loss,''
  Astrophys.\ J.\  {\bf 814}, no. 1, 57 (2015)
  doi:10.1088/0004-637X/814/1/57
  [arXiv:1511.08169 [astro-ph.GA]].
  



%\cite{Hopman:2009gz}
\bibitem{Hopman:2009gz} 
  C.~Hopman,
  %``Binary dynamics near a massive black hole,''
  Astrophys.\ J.\  {\bf 700}, 1933 (2009)
  doi:10.1088/0004-637X/700/2/1933
  [arXiv:0906.0374 [astro-ph.CO]].


%\cite{Wen:2002km}
\bibitem{Wen:2002km} 
  L.~Wen,
  %``On the eccentricity distribution of coalescing black hole binaries driven by the Kozai mechanism in globular clusters,''
  Astrophys.\ J.\  {\bf 598}, 419 (2003)
  doi:10.1086/378794
  [astro-ph/0211492].


%\cite{Kocsis:2011jy}
\bibitem{Kocsis:2011jy} 
  B.~Kocsis and J.~Levin,
  %``Repeated Bursts from Relativistic Scattering of Compact Objects in Galactic Nuclei,''
  Phys.\ Rev.\ D {\bf 85}, 123005 (2012)
  doi:10.1103/PhysRevD.85.123005
  [arXiv:1109.4170 [astro-ph.CO]].


%\cite{Antonini:2013tea}
\bibitem{Antonini:2013tea} 
  F.~Antonini, N.~Murray and S.~Mikkola,
  %``Black hole triple dynamics: breakdown of the orbit average approximation and implications for gravitational wave detections,''
  Astrophys.\ J.\  {\bf 781}, 45 (2014)
  doi:10.1088/0004-637X/781/1/45
  [arXiv:1308.3674 [astro-ph.HE]].

%\cite{VanLandingham:2016ccd}
\bibitem{VanLandingham:2016ccd} 
  J.~H.~VanLandingham, M.~C.~Miller, D.~P.~Hamilton and D.~C.~Richardson,
  %``The Role of the Kozai–lidov Mechanism in Black Hole Binary Mergers in Galactic Centers,''
  Astrophys.\ J.\  {\bf 828}, no. 2, 77 (2016)
  doi:10.3847/0004-637X/828/2/77
  [arXiv:1604.04948 [astro-ph.HE]].

%\cite{Hoang:2019kye}
\bibitem{Hoang:2019kye} 
  B.~M.~Hoang, S.~Naoz, B.~Kocsis, W.~Farr and J.~McIver,
  %``Detecting Supermassive Black Hole–induced Binary Eccentricity Oscillations with LISA,''
  Astrophys.\ J.\  {\bf 875}, no. 2, L31 (2019)
  doi:10.3847/2041-8213/ab14f7
  [arXiv:1903.00134 [astro-ph.HE]].

%\cite{cohn1987}
\bibitem{cohn1987} 
H.~ Cohn and R.~M.~ Kulsrud,
%The stellar distribution around a black hole - Numerical integration of the Fokker-Planck equation, 
1978ApJ...226.1087C
doi:10.1086/156685

%\cite{Pfuhl:2011bk}
\bibitem{Pfuhl:2011bk} 
  O.~Pfuhl {\it et al.},
  %``The star formation history of the Milky Way's Nuclear Star Cluster,''
  Astrophys.\ J.\  {\bf 741}, 108 (2011)
  doi:10.1088/0004-637X/741/2/108
  [arXiv:1110.1633 [astro-ph.GA]].


%\cite{Schodel:2017vjf}
\bibitem{Schodel:2017vjf} 
  R.~Schödel, E.~Gallego-Cano, H.~Dong, F.~Nogueras-Lara, A.~T.~Gallego-Calvente, P.~Amaro-Seoane and H.~Baumgardt,
  %``The distribution of stars around the Milky Way’s central black hole - II. Diffuse light from sub-giants and dwarfs,''
  Astron.\ Astrophys.\  {\bf 609}, A27 (2018)
  doi:10.1051/0004-6361/201730452
  [arXiv:1701.03817 [astro-ph.GA]].
  %%CITATION = doi:10.1051/0004-6361/201730452;%%

%\cite{E. Gallego-Cano}
\bibitem{E. Gallego-Cano} 
E. Gallego-Cano, R. Schödel, H. Dong, F. Nogueras-Lara, A. T. Gallego-Calvente, P. Amaro-Seoane, H. Baumgardt
doi:https:10.1051/0004-6361/201730451
A\&A 609, A26 (2018)
[arxiv:1701.03816]

%\cite{Baumgardt:2017lmw}
\bibitem{Baumgardt:2017lmw} 
  H.~Baumgardt, P.~Amaro-Seoane and R.~Schödel,
  %``The distribution of stars around the Milky Way’s central black hole - III. Comparison with simulations,''
  Astron.\ Astrophys.\  {\bf 609}, A28 (2018)
  doi:10.1051/0004-6361/201730462
  [arXiv:1701.03818 [astro-ph.GA]].


%\cite{Emami:2018taj}
\bibitem{Emami:2018taj} 
  R.~Emami and A.~Loeb,
  %``Formation Redshift of the Massive Black Holes Detected by LIGO,''
  arXiv:1810.09257 [astro-ph.HE].

%\cite{Aharon2016}
\bibitem{Aharon2016}
Danor Aharon,  Alessandra Mastrobuono Battisti, Hagai  B. Perets
%The History of Tidal Disruption Events in Galactic Nuclei
2016ApJ...823..137A
[arXiv:1507.08287]


%\cite{Merritt:2013}
 \bibitem{Merritt:2013}
David, Dynamics and Evolution of Galactic Nuclei, by David Merritt. 
ISBN: 9780691158600. 544 pp. Princeton University Press, 2013
Bibliographic Code:	2013degn.book.....M


%\cite{Bahcall-Wolf-1977}
 \bibitem{Bahcall-Wolf-1977}
% ``THE STAR DISTRIBUTION AROUND A MASSIVE BLACK HOLE IN A GLOBULAR CLUSTER. II. UNEQUAL STAR MASSES*"
Bahcall J. N., Wolf R. A., 1977, ApJ, 216, 883

%\cite{Sedda:2018znc}
\bibitem{Sedda:2018znc} 
  M.~Arca-Sedda and A.~Gualandris,
  %``Gravitational wave sources from inspiralling globular clusters in the Galactic Centre and similar environments,''
  Mon.\ Not.\ Roy.\ Astron.\ Soc.\  {\bf 477}, no. 4, 4423 (2018)
  doi:10.1093/mnras/sty922
  [arXiv:1804.06116 [astro-ph.GA]].

%\cite{Arca-Sedda:2017wea}
\bibitem{Arca-Sedda:2017wea} 
  M.~Arca-Sedda and R.~Capuzzo-Dolcetta,
  %``The MEGaN project II. Gravitational waves from intermediate mass- and binary black holes around a supermassive black hole,''
  Mon.\ Not.\ Roy.\ Astron.\ Soc.\  {\bf 483}, 152 (2019)
  doi:10.1093/mnras/sty3096
  [arXiv:1709.05567 [astro-ph.GA]].
  
  
%\cite{Panamarev:2018bwq}
\bibitem{Panamarev:2018bwq} 
  T.~Panamarev, A.~Just, R.~Spurzem, P.~Berczik, L.~Wang and M.~A.~Sedda,
  %``Direct N-body simulation of the Galactic centre,''
  Mon.\ Not.\ Roy.\ Astron.\ Soc.\  {\bf 484}, 3279 (2019)
  doi:10.1093/mnras/stz208
  [arXiv:1805.02153 [astro-ph.GA]]. 

%\cite{Barack:2003fp}
\bibitem{Barack:2003fp} 
  L.~Barack and C.~Cutler,
  %``LISA capture sources: Approximate waveforms, signal-to-noise ratios, and parameter estimation accuracy,''
  Phys.\ Rev.\ D {\bf 69}, 082005 (2004)
  doi:10.1103/PhysRevD.69.082005
  [gr-qc/0310125].

%\cite{Berry:2012im}
\bibitem{Berry:2012im} 
  C.~P.~L.~Berry and J.~R.~Gair,
  %``Observing the Galaxy's massive black hole with gravitational wave bursts,''
  Mon.\ Not.\ Roy.\ Astron.\ Soc.\  {\bf 429}, 589 (2013)
  doi:10.1093/mnras/sts360
  [arXiv:1210.2778 [astro-ph.HE]].


%\cite{Freitag:2002nm}
\bibitem{Freitag:2002nm} 
  M.~Freitag,
  %``Gravitational waves from stars orbiting the massive black hole at the galactic center,''
  Astrophys.\ J.\  {\bf 583}, L21 (2003)
  doi:10.1086/367813
  [astro-ph/0211209].


%\cite{Gair:2017ynp}
\bibitem{Gair:2017ynp} 
  J.~R.~Gair, S.~Babak, A.~Sesana, P.~Amaro-Seoane, E.~Barausse, C.~P.~L.~Berry, E.~Berti and C.~Sopuerta,
  %``Prospects for observing extreme-mass-ratio inspirals with LISA,''
  J.\ Phys.\ Conf.\ Ser.\  {\bf 840}, no. 1, 012021 (2017)
  doi:10.1088/1742-6596/840/1/012021
  [arXiv:1704.00009 [astro-ph.GA]].
  %%CITATION = doi:10.1088/1742-6596/840/1/012021;%%
  %9 citations counted in INSPIRE as of 19 Apr 2019


%\cite{Emami:2019uty}
\bibitem{Emami:2019uty} 
  R.~Emami and A.~Loeb,
  %``Gravitational Waves from Stellar Mass Black Holes Around SgrA*,''
  arXiv:1903.02579 [astro-ph.HE].

%\cite{Huerta:2015pva}
\bibitem{Huerta:2015pva} 
  E.~A.~Huerta, S.~T.~McWilliams, J.~R.~Gair and S.~R.~Taylor,
  %``Detection of eccentric supermassive black hole binaries with pulsar timing arrays: Signal-to-noise ratio calculations,''
  Phys.\ Rev.\ D {\bf 92}, no. 6, 063010 (2015)
  doi:10.1103/PhysRevD.92.063010
  [arXiv:1504.00928 [gr-qc]].


%\cite{Sesana:2004gf}
\bibitem{Sesana:2004gf} 
 A.~Sesana, F.~Haardt, P.~Madau and M.~Volonteri,
 %``The gravitational wave signal from massive black hole binaries and its contribution to the LISA data stream,''
 Astrophys.\ J.\  {\bf 623}, 23 (2005)
 doi:10.1086/428492
 [astro-ph/0409255].


%\cite{Audley:2017drz}
\bibitem{Audley:2017drz} 
  H.~Audley {\it et al.} [LISA Collaboration],
  %``Laser Interferometer Space Antenna,''
  arXiv:1702.00786 [astro-ph.IM].
		
%\cite{Emami2020}		
\bibitem{Emami2020}		
R.Emami, Eugene Visiliav, et. al. 
To be appread. 
		
		
%\cite{AmaroSeoane:2012tx}
\bibitem{AmaroSeoane:2012tx} 
  P.~Amaro-Seoane,
  %``Relativistic dynamics and extreme mass ratio inspirals,''
  Living Rev.\ Rel.\  {\bf 21}, no. 1, 4 (2018)
  doi:10.1007/s41114-018-0013-8
  [arXiv:1205.5240 [astro-ph.CO]].
		
		
%\cite{Fang:2019dnh}
\bibitem{Fang:2019dnh} 
 X.~Fang, T.~A.~Thompson and C.~M.~Hirata,
 %``The Population of Eccentric Binary Black Holes: Implications for mHz Gravitational Wave Experiments,''
 Astrophys.\ J.\  {\bf 875}, no. 1, 75 (2019)
 doi:10.3847/1538-4357/ab0e6a
 [arXiv:1901.05092 [astro-ph.HE]].		
		
		
		
\end{thebibliography}
\end{document}